%% file: URMC.tex
\documentclass[12pt]{article}

\input{preamble}

\begin{document}

\title{
Nonparametric and Semiparametric Estimation of Upward Rank Mobility Curves\thanks{We thank Le-Yu Chen, Harold D.\ Chiang, Yu-Chin Hsu, Robert P.\ Lieli, Christoph Rothe, Yuya Sasaki, and participants at AMES 2022, EcoSta 2023, SETA 2024, IAAE 2024, as well as seminar participants at Academia Sinica and National Taiwan University for their helpful comments. Tsung-Chih Lai gratefully acknowledges research support from the National Science and Technology Council, Taiwan (NSTC 112-2410-H-194-101-MY2).}
}

\author{
Tsung-Chih Lai\thanks{Department of Economics, National Chung Cheng University, and Department of Finance, National Chung Hsing University.}
\and
Jia-Han Shih\thanks{Department of Applied Mathematics, National Sun Yat-sen University. Corresponding author. Email: \url{jhshih@math.nsysu.edu.tw}}
\and
Yi-Hau Chen\thanks{Institute of Statistical Science, Academia Sinica.}
}

\date{\vspace{-\baselineskip}}

\maketitle

\begin{abstract}
We introduce the \emph{upward rank mobility curve} as a new measure of intergenerational mobility that captures upward movements across the entire parental income distribution. Our approach extends \citet{Bhattacharya2011} by conditioning on a single parental income rank, thereby eliminating aggregation bias. We show that the measure can be characterized solely by the copula of parent and child income, and we propose a nonparametric copula-based estimator with better properties than kernel-based alternatives. For a conditional version of the measure without such a representation, we develop a two-step semiparametric estimator based on distribution regression and establish its asymptotic properties. An application to U.S.\ data reveals that whites exhibit significant upward mobility dominance over blacks among lower-middle-income families.

\vfill


\end{abstract}

\clearpage

\section{Introduction}
\label{sec:intro}

Economists have long been interested in the intergenerational transmission of income, and how to document income mobility across generations has been an active area of research; see \citet{Solon1999,Black2011,Corak2013} for comprehensive reviews and \citet{Deutscher2023} for a recent survey. Among existing measures, perhaps the most widely used is the intergenerational income elasticity \citep[IGE,][]{Solon1992}, defined as the slope from a linear regression of log child income on log parent income. Yet the linearity assumption underlying the IGE may be restrictive, motivating alternative measures such as the rank-rank slope \citep{Chetty2014}, which captures the correlation between children's and parents' income ranks.

While the IGE and rank-rank slope provide useful summaries of persistence, they only indicate the extent of intergenerational mobility rather than its direction. To study upward mobility, researchers have employed transition probabilities across income classes \citep{Formby2004} or the upward rank mobility measure of \citet{Bhattacharya2011}, defined as the probability that a child's relative income rank exceeds that of the parents within a given class. These directional measures have been used extensively \citep[e.g.,][]{Jappelli2006,Dardanoni2012,Hnatkovska2013,Corak2014,Bratberg2017,Richey2018,Millimet2020,Collins2022,Bavaro2023}, but they face two important limitations. First, discretizing the income distribution into classes introduces aggregation bias due to heterogeneity within groups. Second, there is no consensus on how many classes to use, or whether an optimal choice even exists.\footnote{A similar problem arises in estimating the Gini coefficient from grouped data; see, e.g., \citet{Lerman1989,Davies1989}.}

This paper introduces the \emph{upward rank mobility curve}, a new measure that addresses these limitations and facilitates comparisons of upward mobility across the entire parental income distribution. Our measure extends \citet{Bhattacharya2011} by conditioning on a single parental income rank rather than an interval, thereby eliminating aggregation bias. We show that the measure can be expressed solely in terms of the copula of parent and child income---a novel characterization in the literature---and propose a nonparametric estimator based on the empirical Bernstein copula \citep{Sancetta2004}. Compared with the Nadaraya-Watson kernel estimator, our method exhibits improved asymptotic properties. We also provide practical guidance on choosing the smoothing parameter to minimize the asymptotic mean squared error.

Nevertheless, when analyzing differences in upward rank mobility across demographic groups or geographic regions, valid comparisons must rely on ranks derived from the overall income distribution rather than from group-specific ones. In this conditional context, the copula-based representation does not apply. We therefore develop a two-step semiparametric estimator based on distribution regression \citep{Foresi1995,Chernozhukov2013}, establish its asymptotic properties, and examine its finite-sample performance through simulations. Applying these methods to data from the National Longitudinal Survey of Youth, we investigate interracial differences in upward mobility in the United States. The results show significant upward mobility dominance of whites over blacks among lower-middle-income families, corroborating findings from previous studies \citep{Bhattacharya2011,Mazumder2014,Fox2016,Chetty2019}.

A large body of research emphasizes that intergenerational mobility can be measured in multiple ways, with distinct metrics capturing different aspects of the transmission process. As highlighted by \citet{Deutscher2023}, an important distinction is between \emph{global} and \emph{local} measures. Global measures, such as the IGE and the rank-rank slope, summarize the joint distribution of parent and child income with a single parameter. Local measures, by contrast, focus on specific parts of the distribution and reveal heterogeneous patterns. Transition probabilities, conditional expected ranks, and directional rank mobility are particularly useful for detecting barriers to upward mobility at the bottom (poverty traps) or persistence at the top (entrenched privilege); see, e.g., \citet{Corak2020}. Such measures have proven central to understanding nonlinearities in mobility and their links to mechanisms such as credit constraints or neighborhood effects \citep{Grawe2004,Durlauf2004}.

Another key distinction concerns whether mobility is measured in \emph{relative} or \emph{absolute} terms. Relative measures, including the IGE, the rank-rank slope, and the proposed unconditional upward rank mobility curve, compare children's income levels or ranks directly with those of their parents. In contrast, absolute measures assess mobility against an external benchmark, which may be defined by fixed standards of living \citep{Chetty2017}, or by income ranks derived from the overall population for cross-group comparisons. Our proposed conditional upward rank mobility curve falls into this category and is particularly well suited for the latter application.

Because relative and absolute mobility need not coincide, it is important to consider both perspectives. Indeed, \citet{Deutscher2023} show that correlations between them are generally weak, implying that no single metric provides a complete picture of intergenerational mobility. Against this backdrop, the contribution of this paper is to introduce both unconditional and conditional mobility curves, thereby extending the set of directional local measures to encompass both relative and absolute dimensions.

Our work is related to \citet{Callaway2024}, who study the identification and estimation of intergenerational mobility parameters under two-sided measurement error. They show that a broad class of parameters (including transition matrices, upward mobility measures of \citet{Bhattacharya2011}, and rank-rank slopes) can be characterized by the copula of parent and child income as well, and they propose a multi-step semiparametric estimator based on quantile regression. By contrast, our contribution is twofold. First, we introduce upward rank mobility curves that eliminate measurement error arising from misclassification into incorrect transition cells (see \cite{Millimet2020} for a related argument). Second, we exploit the copula representation to develop a nonparametric estimator in the unconditional case. Thus, while their work addresses robustness to measurement error, ours expands the methodological toolkit for analyzing upward rank mobility.

Our work is also related to \citet{Chernozhukov2025}, who introduce bivariate distribution regression as a flexible framework for modeling conditional joint distributions and decomposing intergenerational mobility. In this conditional setting, however, our focus is on diagnosing mobility dominance between groups (e.g., racial disparities) that cannot be inferred from conditional joint distributions alone. As noted earlier, such comparisons require unconditional income ranks of parents and children, which are typically unobserved and must be estimated. We therefore establish the weak convergence of the conditional mobility curve estimator accounting for these estimation effects, and use it to conduct uniform inference across the entire parental income distribution.

The rest of the paper is organized as follows. \cref{sec:urmc} formally introduces the upward rank mobility curve and the nonparametric copula-based estimator. \cref{sec:cond} extends the analysis to the conditional measure and develops a semiparametric distribution regression procedure, along with its asymptotic properties. \cref{sec:simu} and \cref{sec:empi} present the simulation and empirical studies, respectively, and \cref{sec:conc} concludes.

\section{Upward Rank Mobility Curves}
\label{sec:urmc}

\subsection{Definition}

Let $Y_0$ and $Y_1$ denote parental and child income, with marginal cumulative distribution functions (CDFs) $F_0$ and $F_1$, respectively. For families whose (normalized) parental income rank lies within an interval, $F_0(Y_0)\in[s_1,s_2]$, \cite{Bhattacharya2011} propose a measure of upward rank mobility defined as the probability that the child's income rank, $F_1(Y_1)$, exceeds that of the parent by at least $\tau\in[0,1-s_2]$ for $0<s_1<s_2<1$:
\begin{equation}
\label{urm}
\upsilon(\tau,s_1,s_2)\equiv\Prob(F_1(Y_1)>F_0(Y_0)+\tau|s_1\leq F_0(Y_0)\leq s_2).
\end{equation}
Although widely used, this measure is susceptible to aggregation bias arising from heterogeneity among families within $[s_1,s_2]$. In particular, it is unclear whether observed upward mobility originates mainly from families at the lower or upper end of the interval. This bias is further compounded by the common practice of dividing income into only a few categories (i.e., using relatively coarse intervals).

In this paper, we introduce the \emph{upward rank mobility curve}, which conditions on an exact parental income rank rather than on an interval:
\begin{equation}
\label{urmc}
u(\tau,s)\equiv\Prob(F_1(Y_1)>F_0(Y_0)+\tau|F_0(Y_0)=s),
\end{equation}
where $0\leq\tau<1-s$ and $0<s<1$. Clearly, $u(\tau,s)=\upsilon(\tau,s,s+t)$ as $t\to 0$. This refinement eliminates aggregation bias and facilitates comparisons of intergenerational upward mobility in terms of stochastic dominance \citep{Fields2002} and stochastic monotonicity \citep{Lee2009} across the parental income distribution. For example, a formal definition of upward mobility dominance is provided in \cref{sec:cond}.

As pointed out by \citet{Bhattacharya2011}, conditioning on $F_0(Y_0)=s$ typically requires averaging over a bandwidth around $s$ when kernel smoothing is applied. However, selecting an optimal bandwidth that balances bias and variance is challenging as it depends on the unknown marginals $F_0$ and $F_1$. Moreover, the Nadaraya-Watson estimator is well known to suffer from boundary effects, which lead to larger bias when estimating $u(\tau,s)$ near $s=0$ or $s=1$. To overcome these limitations, we propose a novel nonparametric method that avoids estimating the marginal distributions. We also derive an asymptotic expression for the mean squared error, valid uniformly over interior and boundary points, which in turn yields the optimal smoothing parameter for our estimator.

\subsection{Empirical Bernstein Copula-Based Estimators}
\label{sec:ebc}

We first show that the mobility measure defined in \cref{urmc} can be expressed solely in terms of the copula of $(Y_0,Y_1)$, complementing the findings of \citet{Callaway2024} that connect relative intergenerational mobility to the copula literature. This result also parallels \citet{Chetty2017}, who link absolute intergenerational mobility to the copula through decomposition.
\begin{pro}
\label{pro:urmc}
Let $\partial_0C\equiv\partial C(u_0,u_1)/\partial u_0$ denote the first partial derivative of the copula $C$ of $(Y_0,Y_1)$. If $F_0$ and $F_1$ are continuous and strictly increasing, we have 
\[
u(\tau,s)=1-\partial_0C(s,s+\tau)
\]
for almost all $\tau\in[0,1-s)$ and $s\in(0,1)$ with respect to Lebesgue measure.
\end{pro}

From \cref{pro:urmc} (see \cref{sec:proof} for the proof), the problem now reduces to estimating the copula derivative. Our nonparametric estimator builds on the work of \cite{Janssen2016}, who employ Bernstein estimation for the first-order derivative of a copula. Specifically, let $\{(Y_{i0},Y_{i1}):i=1,\dotsc,n\}$ be a random sample of $(Y_0,Y_1)$. Denote $R_{ij}=\sum_{k=1}^n\1\{Y_{kj}\leq Y_{ij}\}$ as the rank of $Y_{ij}$ among $Y_{1j},\dotsc,Y_{nj}$ for $i=1,\dotsc,n$ and $j=0,1$, with $\1\{\cdot\}$ being the indicator function. The Bernstein estimator of \cite{Janssen2016} is essentially the partial derivative of the empirical Bernstein copula, $C_{m,n}$, introduced by \cite{Sancetta2004}:
\begin{align}
\label{ebc-partial}
\widehat{\partial_0C}_{m,n}(u_0,u_1)&=
\frac{\partial}{\partial u_0}C_{m,n}(u_0,u_1)\notag\\
&=\frac{\partial}{\partial u_0}\sum_{k=0}^m\sum_{\ell=0}^m C_n\left(\frac{k}{m},\frac{\ell}{m}\right)P_{m,k}(u_0)P_{m,\ell}(u_1)\notag\\
&=m\sum_{k=0}^{m-1}\sum_{\ell=0}^m\left(C_n\left(\frac{k+1}{m},\frac{\ell}{m}\right)-C_n\left(\frac{k}{m},\frac{\ell}{m}\right)\right)P_{m-1,k}(u_0)P_{m,\ell}(u_1),
\end{align}
where $m\in\mathbb{N}$ is the order of the Bernstein polynomial satisfying $m\to\infty$ and $m/n\to0$ as $n\to\infty$, $C_n(u_0,u_1)=n^{-1}\sum_{i=1}^n\1\left\{R_{i0}/n\leq u_0,R_{i1}/n\leq u_1\right\}$ is the rank-based empirical copula, and $P_{m,k}(u)={m \choose k}u^k(1-u)^{m-k}$ is the binomial probability for $k=0,1,\dotsc,m$. Note that the definition of $C_n$ here differs slightly from that in \cite{Janssen2016}, but the discrepancy between these empirical copula variants is at most $2/n$; see \cite{Fermanian2004}.

According to \cref{pro:urmc,ebc-partial}, the empirical Bernstein copula-based (EBC) estimator of $u(\tau,s)$ is defined as
\begin{equation}
\label{urmc-ebc}
\widehat{u}^{\text{EBC}}_{m,n}(\tau,s)\equiv1-\widehat{\partial_0C}_{m,n}(s,s+\tau).
\end{equation}
The remaining task is to choose the order $m$. Assuming $C$ has third-order partial derivatives that are Lipschitz continuous in $(0,1)^2$, the results of \citet{Swanepoel2013} yield the asymptotic mean squared error of $\widehat{u}^{\text{EBC}}_{m,n}(\tau,s)$, uniformly in $\tau$ and $s$:
\begin{align*}
\AMSE(\widehat{u}^{\text{EBC}}_{m,n}(\tau,s))
&=\frac{1}{4m^2}b^2(s,s+\tau)+\frac{m^{1/2}}{n}\sigma^2(s,s+\tau),
\end{align*}
where
\begin{align*}
b(u_0,u_1)&\equiv(1-2u_0)\frac{\partial^2}{\partial u_0^2}C(u_0,u_1)+u_0(1-u_0)\frac{\partial^3}{\partial u_0^3}C(u_0,u_1)+u_1(1-u_1)\frac{\partial^3}{\partial u_0\partial u_1^2}C(u_0,u_1),\\
\sigma^2(u_0,u_1)&\equiv\frac{\partial_0C(u_0,u_1)(1-\partial_0C(u_0,u_1))}{2\sqrt{\pi u_0(1-u_0)}}. 
\end{align*}
Therefore, the optimal order that minimizes the asymptotic mean squared error is
\begin{equation}
\label{m-optimal}
m^*=m^*(\tau,s)=\left\lceil\left(\frac{b^2(s,s+\tau)}{\sigma^2(s,s+\tau)}\right)^{2/5}n^{2/5}\right\rceil,
\end{equation}
where $\lceil x\rceil$ denotes the smallest integer not smaller than $x$.\footnote{When $b(s,s+\tau)=0$, we simply set $m^*=2$.} In practice, however, $m^*$ is difficult to implement because it depends on third-order partial derivatives of the copula, which are typically unknown. Moreover, since $m^*$ varies with $(\tau,s)$, it is impractical to use different orders for each point on the curve. In simulations reported in \cref{sec:simu}, we find that setting $m=n^{1/2}$ globally performs as well as, or even better than, the pointwise optimal $m^*$ in finite samples.

\begin{rmk}
As summarized in \citet{Janssen2016}, the Bernstein estimator in \cref{ebc-partial} for $(u_0,u_1)\in(0,1)^2$ exhibits superior asymptotic properties compared with the Nadaraya-Watson estimator. For comparison, we set $h=m^{-1}$ as the bandwidth following \citet{Sancetta2004}. The asymptotic variance of the Bernstein estimator is of order $O(m^{1/2}/n)$, whereas that of the Nadaraya-Watson estimator is $O(m/n)$. With respect to bias, the Bernstein estimator achieves an order of $O(m^{-1})$ at both interior and boundary points (and is therefore free from boundary effects), while the Nadaraya-Watson estimator attains $O(m^{-2})$ at interior points and only $O(m^{-1})$ at boundary points. Consequently, the optimal mean squared error of the Bernstein estimator is $O(n^{-4/5})$ uniformly at interior and boundary points, whereas for the Nadaraya-Watson estimator it is $O(n^{-4/5})$ at interior points but deteriorates to $O(n^{-2/3})$ at boundary points.
\end{rmk}

\begin{rmk}
Instead of minimizing the asymptotic mean squared error, one may alternatively minimize the asymptotic bias by setting $m=n$, i.e., the largest possible order, noting that the asymptotic variance of the Bernstein estimator is of smaller magnitude. Interestingly, such undersmoothing renders the estimator in \cref{ebc-partial} identical to the partial derivative of the empirical beta copula, $C_n^\beta$, developed by \citet{Segers2017}:
\begin{align*}
\widehat{\partial_0C}_{n,n}(u_0,u_1)
&=\frac{\partial}{\partial u_0}C^\beta_n(u_0,u_1)\\
&=\frac{\partial}{\partial u_0}\frac{1}{n}\sum_{i=1}^nF_{n,R_{i0}}(u_0)F_{n,R_{i1}}(u_1)\\
&=\frac{1}{n}\sum_{i=1}^nf_{n,R_{i0}}(u_0)F_{n,R_{i1}}(u_1),
\end{align*}
where $F_{n,r}(u)=\sum_{s=r}^n{n\choose s}u^s(1-u)^{n-s}$ is the CDF of the beta distribution $\mathcal{B}(r,n+1-r)$ for $r=1,\dotsc,n$, and $f_{n,r}$ is the corresponding probability density function. A key advantage of this estimator is that it eliminates the need to select a smoothing parameter. As shown in \cref{sec:simu}, the empirical beta copula-based estimator of $u(\tau,s)$,
\begin{equation}
\label{urmc-beta}
\widehat{u}^\beta_n(\tau,s)\equiv1-\widehat{\partial_0C}_{n,n}(s,s+\tau),
\end{equation}
substantially outperforms competing estimators in terms of bias.
\end{rmk}

\section{Conditional Upward Rank Mobility Curves}
\label{sec:cond}

While $u(\tau,s)$ effectively measures upward rank mobility across the entire parental income distribution, examining its conditional counterpart given covariates $X$ can provide further insight into mobility differences across demographic groups or geographic regions. This section extends our analysis to such a conditional measure. However, unlike in \cref{pro:urmc}, no analogous copula representation is available when ranks are defined with respect to the overall population rather than the group under study, rendering the method in \cref{sec:urmc} inapplicable. We therefore propose an alternative two-step semiparametric estimator based on distribution regression and establish its uniform asymptotic properties, which are useful for testing dominance in upward rank mobility.

\subsection{Definition}

Our definition of the \emph{conditional upward rank mobility curve} is a direct modification of \citet{Bhattacharya2011}:
\begin{equation}
\label{cond}
u_c(x;\tau,s)\equiv\Prob(F_1(Y_1)>F_0(Y_0)+\tau|F_0(Y_0)=s,X=x).
\end{equation}
where $0\leq\tau<1-s$ and $0<s<1$. Importantly, $F_0(Y_0)$ and $F_1(Y_1)$ here still denote the \emph{unconditional} income ranks of parents and children. As noted by \citet{Deutscher2023}, such a conditional measure behaves like an absolute measure, since the ranks are defined relative to a fixed external benchmark. This feature enables meaningful between-group comparisons, rather than restricting attention to mobility within the group defined by $X=x$. Accordingly, following \cite{Aaberge2014}, we define upward mobility dominance between two groups as follows:
\begin{defn}
\label{defn}
The $x_1$-group is said to exhibit upward mobility dominance over the $x_2$-group on $\mathcal{S}\subseteq(0,1)$ if
\[
u_c(x_1;\tau,s)\geq u_c(x_2;\tau,s)\quad\text{for all $s\in\mathcal{S}$}
\]
and the inequality holds strictly for some $s\in\mathcal{S}$.
\end{defn}

Unlike the representation in \cref{pro:urmc}, \cref{cond} does not admit a straightforward copula-based characterization. A more natural conditional analogue is instead given by
\begin{align}
\widetilde{u}_c(x;\tau,s)
&\equiv\Prob(F_{1|X}(Y_1|x)>F_{0|X}(Y_0|x)+\tau|F_{0|X}(Y_0|x)=s,\,X=x)\label{cond-alter}\\
&=1-\partial_0C_x(s,s+\tau),\notag
\end{align}
where $F_{0|X}$ and $F_{1|X}$ denote the conditional income distributions of parents and children given $X$, and $C_x$ is the conditional copula satisfying $\Prob(Y_0\leq y_0,Y_1\leq y_1|X=x)=C_x(F_{0|X}(y_0|x),F_{1|X}(y_1|x))$. The distinction between \cref{cond} and \cref{cond-alter} is clear in applications. If $X$ denotes race, \cref{cond} captures \emph{interracial} differences in upward mobility relative to the overall population, while \cref{cond-alter} reflects \emph{intraracial} differences within the income distribution of a particular racial group. Since our interest lies in between-group comparisons, we focus on \cref{cond} in the remainder of the paper.

\subsection{Distribution Regression-Based Estimators}
\label{sec:dr}

To estimate \cref{cond}, note that if $F_0$ and $F_1$ are continuous and strictly increasing, the conditional measure can be rewritten as
\begin{equation}
\label{cond-ccdf}
u_c(x;\tau,s)=1-F_{1|0,X}(Q_1(s+\tau)|Q_0(s),x),
\end{equation}
where $F_{1|0,X}$ is the conditional CDF of $Y_1$ given $Y_0$ and $X$, and $Q_j=F_j^{-1}$ is the quantile function of $Y_j$ for $j=0,1$. Estimation therefore proceeds in two steps: (i) construct an estimator of $F_{1|0,X}(y_1|y_0,x)$; and (ii) evaluate $y_1$ and $y_0$ at the sample counterparts of $Q_1(s+\tau)$ and $Q_0(s)$, respectively.

Our first step follows the distribution regression developed by \cite{Foresi1995,Chernozhukov2013}:
\begin{equation}
\label{ccdf-dr}
\widehat{F}_{1|0,X}(y_1|y_0,x)=\Lambda(P(y_0,x)^\top\widehat\theta(y_1)),
\end{equation}
where $\Lambda$ is a link function, $P(Y_0,X)$ is a vector of polynomials of $Y_0$ and $X$, and $\widehat\theta(y_1)$ is the parameter vector indexed by $y_1$ that maximizes the likelihood:
\[
\sum_{i=1}^n(\1\{Y_{i1}\leq y_1\}\ln\Lambda(P(Y_{i0},X_i)^\top\theta(y_1))+\1\{Y_{i1}>y_1\}\ln(1-\Lambda(P(Y_{i0},X_i)^\top\theta(y_1)))).
\]
This semiparametric approach provides a flexible way to model the conditional distribution of children's income given parental income and observed covariates. Rather than assuming a fully parametric model, it uses a series of binary regressions at different thresholds, which allows the conditional distribution to be approximated arbitrarily well; see \citet{Chernozhukov2013} for details. This approach enables us to evaluate upward rank mobility at specific parental ranks while accommodating rich forms of heterogeneity across groups.

Combining \cref{cond-ccdf,ccdf-dr}, the distribution regression-based (DR) estimator of $u_c(x;\tau,s)$ is defined as
\begin{equation}
\label{cond-dr}
\widehat u^\text{DR}_c(x;\tau,s)\equiv1-\widehat F_{1|0,X}(\widehat Q_1(s+\tau)|\widehat Q_0(s),x),
\end{equation}
where $\widehat{Q}_j(p)=\inf\{y\in\R:\widehat F_j(y)\geq p\}$ is the empirical quantile function of $Y_j$ for $j=0,1$, with $\widehat F_j(y)=n^{-1}\sum_{i=1}^n\1\{Y_{ij}\leq y\}$ the corresponding empirical CDF. For comparison with the nonparametric EBC estimator in \cref{urmc-ebc}, we also define the DR estimator of the unconditional mobility curve $u(\tau,s)$ as
\begin{equation}
\label{urmc-dr}
\widehat u^\text{DR}(\tau,s)\equiv1-\widehat F_{1|0}(\widehat Q_1(s+\tau)|\widehat Q_0(s)),
\end{equation}
where $\widehat F_{1|0}$ is constructed analogously to \cref{ccdf-dr}.

\subsection{Asymptotic Properties}

We now establish the weak convergence of the DR estimator defined in \cref{urmc-dr}. Similar results hold for the conditional estimator in \cref{cond-dr}, but are omitted for brevity. In both cases, however, it is important to account for the estimation effects introduced by the empirical quantile functions $\widehat Q_0$ and $\widehat Q_1$ when deriving the limiting processes. The regularity conditions below are adapted from \citet{Chernozhukov2013}.

\begin{asm}
\label{asm:dr}
Denote $\Y_0\subseteq\R$ and $\Y_1\subseteq\R$ as the supports of $Y_0$ and $Y_1$. Suppose that
\begin{enumerate}[(i)]
\item
The support $\Y_0\times\Y_1$ is a compact subset of $\R^2$.
\item
The distribution function $F_j$ is uniformly continuous on $\Y_j$ for $j=0,1$.
\item
The density function $f_j$ is bounded away from zero uniformly on $\Y_j$ for $j=0,1$.
\item
The conditional density function $f_{1|0}$ is uniformly continuous and bounded on $\Y_1\times\Y_0$.
\item
$\E\|P(Y_0)\|^2<\infty$ and the minimum eigenvalue of
\[
H(y_1)\equiv\E\left(\frac{\lambda^2(P(Y_0)^\top\theta(y_1))}{\Lambda(P(Y_0)^\top\theta(y_1))(1-\Lambda(P(Y_0)^\top\theta(y_1)))}P(Y_0)P(Y_0)^\top\right)
\]
is bounded away from zero uniformly on $\Y_1$, where $\lambda$ is the derivative of $\Lambda$.
\end{enumerate}
\end{asm}

\begin{thm}
\label{thm:weak}
Suppose $F_{1|0}(y_1|y_0)=\Lambda(P(y_0)^\top\theta(y_1))$ is correctly specified for all $y_1\in\Y_1$ and $y_0\in\Y_0$, and \cref{asm:dr} is satisfied. Then,
\[
\sqrt{n}\left(\widehat u^\text{\upshape DR}(\tau,s)-u(\tau,s)\right)\Rightarrow\Psi(\tau,s),
\]
where $\Rightarrow$ denotes weak convergence and $\Psi(\tau,s)$ is a zero-mean Gaussian process with covariance function generated by the influence function
\begin{align*}
\psi(\tau,s,Y_1,Y_0)&=-\left\{
\lambda(P(Q_0(s))^\top\theta(Q_1(s+\tau)))P(Q_0(s))^\top\psi_\theta(Q_1(s+\tau),Y_1,Y_0)\right.\\
&\quad+\lambda(P(Q_0(s))^\top\theta(Q_1(s+\tau)))P(Q_0(s))^\top\theta'(Q_1(s+\tau))\psi_1(\tau,s,Y_1)\\
&\left.\quad+\lambda(P(Q_0(s))^\top\theta(Q_1(s+\tau)))P'(Q_0(s))^\top\theta(Q_1(s+\tau))\psi_0(s,Y_0)\right\},
\end{align*}
where $\theta'(\cdot)$ and $P'(\cdot)$ are $(p+1)$-dimensional vectors consisting of elementwise derivatives of $\theta(\cdot)$ and $P(\cdot)$, respectively, and
\begin{align*}
\psi_\theta(y_1,Y_1,Y_0)&=H^{-1}(y_1)\frac{\1\{Y_1\leq y_1\}-\Lambda(P(Y_0)^\top\theta(y_1))}{\Lambda(P(Y_0)^\top\theta(y_1))(1-\Lambda(P(Y_0)^\top\theta(y_1)))}\lambda(P(Y_0)^\top\theta(y_1))P(Y_0),\\
\psi_1(\tau,s,Y_1)&=\frac{(s+\tau)-\1\{Y_1\leq Q_1(s+\tau)\}}{f_1(Q_1(s+\tau))},\\
\psi_0(s,Y_0)&=\frac{s-\1\{Y_0\leq Q_0(s)\}}{f_0(Q_0(s))}.
\end{align*}
\end{thm}

\cref{thm:weak} shows that the estimator in \cref{urmc-dr} converges weakly to a zero-mean Gaussian process at the parametric rate. However, since the influence function in \cref{thm:weak} depends on unknown nuisance functions and is therefore non-pivotal, we employ the empirical bootstrap, following \citet{Chernozhukov2013}, to conduct uniform inference for the entire curve. For example, the bootstrap $(1-\alpha)$ uniform confidence band for $u(\tau,s)$ with fixed $\tau$ is given by
\[
\left[\widehat{u}^{\text{DR}}(\tau,s)-c_{1-\alpha}^B\widehat\sigma^B(\tau,s),~\widehat{u}^\text{DR}(\tau,s)+c_{1-\alpha}^B\widehat\sigma^B(\tau,s)\right],
\]
where $c_{1-\alpha}^B$ is the $(1-\alpha)$ quantile of $\{\sup_{0<s<1}|(\widehat u_b^\text{DR}(\tau,s)-\widehat u^\text{DR}(\tau,s))/\widehat{\sigma}^B(\tau,s)|\}_{b=1}^B$ based on $B$ bootstrap samples. Here, $\widehat u_b^\text{DR}(\tau,s)$ denotes the bootstrap version of $\widehat u^\text{DR}(\tau,s)$ for $b=1,\dotsc,B$, and $\widehat{\sigma}^B(\tau,s)$ is the pointwise bootstrap standard deviation of $\widehat u^\text{DR}(\tau,s)$. An analogous procedure can be applied to test upward mobility dominance in \cref{defn}, through the details are omitted for brevity.

\section{Simulation Study}
\label{sec:simu}

We examine the finite-sample performance of the EBC estimator in \cref{urmc-ebc} and the DR estimator in \cref{urmc-dr}. The data $\{(Y_{i0},Y_{i1})\}_{i=1}^n$ are generated from the following four copulas with standard Gaussian marginal distributions: 
\begin{enumerate}
\item
\textbf{Gaussian copula:} $C_\theta(u_0,u_1) = \Phi_\theta(\Phi^{-1}(u_0),\Phi^{-1}(u_1))$ with $\theta \in (-1,1)$, where $\Phi_\theta$ is the bivariate standard Gaussian CDF with correlation coefficient $\theta$, and $\Phi$ is the univariate standard Gaussian CDF.
\item
\textbf{Clayton copula:} $C_\theta(u_0,u_1) = (u_0^{-\theta}+u_1^{-\theta}-1)^{-1/\theta}$ with $\theta \in (0,\infty)$.
\item
\textbf{Gumbel copula:} $C_\theta(u_0,u_1) = \exp(-((-\ln u_0)^\theta+(-\ln u_1)^\theta)^{1/\theta})$ with $\theta \in [1,\infty)$.
\item 
\textbf{Independence copula:} $C(u_0,u_1) = u_0 u_1$.
\end{enumerate}
For the Gaussian, Clayton, and Gumbel copulas, the parameter $\theta$ is calibrated so that Kendall's tau satisfies $\tau_K\in\{1/3,1/2,2/3\}$, corresponding to weak, moderate, and strong dependence between $Y_0$ and $Y_1$, respectively. Under the independence copula, we have $\tau_K=0$.

For the EBC estimator, we consider three specifications for the order of the Bernstein polynomial $m$:
(i) $\widehat u^\text{EBC}_{m^*,n}$, where $m^*$ is the optimal order given in \cref{m-optimal};
(ii) $\widehat u^\text{EBC}_{n^{1/2},n}$, which uses the intermediate order $n^{1/2}$;
(iii) $\widehat u^\text{EBC}_{n,n} = \widehat u^\beta$, which sets $m=n$ and yields the empirical beta copula in \cref{urmc-beta}.
The optimal order $m^*=m^*(\tau,s)$ is computed pointwise under the true copula. Although infeasible in practice, it is included here for benchmarking. For the DR estimator, we examine three combinations of the link function $\Lambda$ and the polynomial $P(Y_0)$:
(i) $\widehat u^\text{DR}_\text{Probit}$, using the probit link with $P(Y_0)=Y_0$;
(ii) $\widehat u^\text{DR}_\text{Logit}$, using the logit link with $P(Y_0)=Y_0$;
(iii) $\widehat u^\text{DR}_\text{Logit,2}$, using the logit link with $P(Y_0)=[Y_0~Y_0^2]^\top$.
While all these specifications are initially misspecified, we will show that their performance improves substantially as the polynomial order increases.

\begin{table}
\setlength\extrarowheight{3pt}
\begin{threeparttable}
\centering
\caption{Simulation results under the Gaussian copula.\label{table:simu1}}
\begin{tabular*}{\textwidth}{@{\extracolsep{\fill}}ccrrrrrr}
\toprule
Kendall's Tau	&	URMC	&	\multicolumn{3}{c}{RISB}	&	\multicolumn{3}{c}{RIMSE}	\\
\cmidrule(lr){3-5}	\cmidrule(lr){6-8}
(Copula Parameter)	&	Estimator	&	\multicolumn{1}{c}{100}	&	\multicolumn{1}{c}{200}	&	\multicolumn{1}{c}{400}	&	\multicolumn{1}{c}{100}	&	\multicolumn{1}{c}{200}	&	\multicolumn{1}{c}{400}	\\
\midrule
	                    &	$\widehat u^\text{EBC}_{m^*,n}$         &	3.597	&	2.691	&	1.946	&	3.934	&	3.014	&	2.281	\\
	                    &	$\widehat u^\text{EBC}_{n^{1 / 2},n}$	&	2.118	&	1.455	&	1.103	&	3.890	&	3.300	&	2.710	\\
$\tau_K=\frac{1}{3}$	&	$\widehat u^\beta$	                    &	0.398	&	0.340	&	0.266	&	10.094	&	9.017	&	7.918	\\
($\theta=0.5$)	        &	$\widehat u^\text{DR}_\text{Probit}$	&	0.228	&	0.156	&	0.103	&	3.594	&	2.544	&	1.841	\\
	                    &	$\widehat u^\text{DR}_\text{Logit}$	    &	0.455	&	0.357	&	0.339	&	3.509	&	2.501	&	1.833	\\
	                    &	$\widehat u^\text{DR}_\text{Logit,2}$	&	0.604	&	0.364	&	0.253	&	6.300	&	3.862	&	2.585	\\
\midrule
	                    &	$\widehat u^\text{EBC}_{m^*,n}$         &	4.610	&	3.347	&	2.386	&	4.883	&	3.623	&	2.682	\\
	                    &	$\widehat u^\text{EBC}_{n^{1/2},n}$	    &	3.686	&	2.610	&	2.010	&	4.652	&	3.705	&	3.010	\\
$\tau_K=\frac{1}{2}$	&	$\widehat u^\beta$	                    &	0.502	&	0.367	&	0.305	&	10.059	&	9.146	&	8.140	\\
($\theta=0.707$)	    &	$\widehat u^\text{DR}_\text{Probit}$	&	0.790	&	0.152	&	0.076	&	4.256	&	3.036	&	2.191	\\
	                    &	$\widehat u^\text{DR}_\text{Logit}$	    &	1.221	&	0.713	&	0.626	&	4.341	&	3.110	&	2.275	\\
	                    &	$\widehat u^\text{DR}_\text{Logit,2}$	&	0.966	&	0.630	&	0.486	&	7.623	&	4.761	&	3.122	\\
\midrule
	                    &	$\widehat u^\text{EBC}_{m^*,n}$         &	5.292	&	3.903	&	2.834	&	5.477	&	4.090	&	3.032	\\
	                    &	$\widehat u^\text{EBC}_{n^{1/2},n}$	    &	5.613	&	4.115	&	3.228	&	5.982	&	4.550	&	3.661	\\
$\tau_K=\frac{2}{3}$	&	$\widehat u^\beta$	                    &	0.776	&	0.372	&	0.307	&	9.085	&	8.651	&	8.007	\\
($\theta=0.866$)	    &	$\widehat u^\text{DR}_\text{Probit}$	&	2.393	&	0.922	&	0.285	&	5.916	&	3.976	&	2.773	\\
	                    &	$\widehat u^\text{DR}_\text{Logit}$	    &	2.888	&	1.505	&	0.980	&	6.218	&	4.253	&	3.020	\\
	                    &	$\widehat u^\text{DR}_\text{Logit,2}$	&	1.081	&	0.596	&	0.528	&	9.078	&	5.704	&	3.809	\\
\bottomrule
\end{tabular*}
\begin{tablenotes}
\item 
Note: The results are obtained from 1,000 simulations. See texts for more details.
\end{tablenotes}
\end{threeparttable}
\end{table}

\begin{table}
\setlength\extrarowheight{3pt}
\begin{threeparttable}
\centering
\caption{Simulation results under the Clayton copula.\label{table:simu2}}
\begin{tabular*}{\textwidth}{@{\extracolsep{\fill}}ccrrrrrr}
\toprule
Kendall's Tau	&	URMC	&	\multicolumn{3}{c}{RISB}	&	\multicolumn{3}{c}{RIMSE}	\\
\cmidrule(lr){3-5}	\cmidrule(lr){6-8}
(Copula Parameter)	&	Estimator	&	\multicolumn{1}{c}{100}	&	\multicolumn{1}{c}{200}	&	\multicolumn{1}{c}{400}	&	\multicolumn{1}{c}{100}	&	\multicolumn{1}{c}{200}	&	\multicolumn{1}{c}{400}	\\
\midrule
	                    &	$\widehat u^\text{EBC}_{m^*,n}$         &	6.318	&	5.648	&	5.017	&	6.613	&	5.893	&	5.230	\\
	                    &	$\widehat u^\text{EBC}_{n^{1/2},n}$     &	3.443	&	2.519	&	1.999	&	4.695	&	3.821	&	3.144	\\
$\tau_K=\frac{1}{3}$    &	$\widehat u^\beta$	                    &	0.636	&	0.289	&	0.295	&	9.803	&	8.823	&	7.797	\\
($\theta=1$)         	&	$\widehat u^\text{DR}_\text{Probit}$	&	4.164	&	3.983	&	3.953	&	5.744	&	4.925	&	4.474	\\
             	        &	$\widehat u^\text{DR}_\text{Logit}$	    &	4.022	&	3.813	&	3.771	&	5.548	&	4.720	&	4.259	\\
	                    &	$\widehat u^\text{DR}_\text{Logit,2}$	&	1.029	&	0.589	&	0.425	&	6.911	&	4.412	&	2.877	\\
\midrule
	                    &	$\widehat u^\text{EBC}_{m^*,n}$         &	6.282	&	5.427	&	4.678	&	6.761	&	5.816	&	5.010	\\
	                    &	$\widehat u^\text{EBC}_{n^{1/2},n}$	    &	5.097	&	3.806	&	3.048	&	5.801	&	4.566	&	3.734	\\
$\tau_K=\frac{1}{2}$    &	$\widehat u^\beta$	                    &	0.879	&	0.406	&	0.329	&	9.534	&	8.702	&	7.748	\\
($\theta=2$)            &	$\widehat u^\text{DR}_\text{Probit}$	&	6.019	&	5.551	&	5.362	&	7.756	&	6.598	&	5.931	\\
                       	&	$\widehat u^\text{DR}_\text{Logit}$	    &	5.803	&	5.220	&	4.981	&	7.597	&	6.340	&	5.602	\\
	                    &	$\widehat u^\text{DR}_\text{Logit,2}$	&	2.232	&	1.375	&	0.841	&	8.308	&	5.460	&	3.623	\\
\midrule
	                    &	$\widehat u^\text{EBC}_{m^*,n}$         &	6.181	&	5.239	&	4.401	&	6.769	&	5.712	&	4.814	\\
	                    &	$\widehat u^\text{EBC}_{n^{1/2},n}$	    &	6.665	&	5.075	&	4.137	&	6.982	&	5.442	&	4.479	\\
$\tau_K=\frac{2}{3}$	&	$\widehat u^\beta$	                    &	1.229	&	0.603	&	0.384	&	8.632	&	8.119	&	7.344	\\
($\theta=4$)	        &	$\widehat u^\text{DR}_\text{Probit}$	&	7.985	&	6.817	&	6.235	&	10.532	&	8.435	&	7.174	\\
	                    &	$\widehat u^\text{DR}_\text{Logit}$	    &	7.872	&	6.517	&	5.803	&	10.480	&	8.251	&	6.858	\\
                    	&	$\widehat u^\text{DR}_\text{Logit,2}$	&	5.200	&	3.535	&	2.222	&	11.235	&	7.879	&	5.522	\\
\bottomrule
\end{tabular*}
\begin{tablenotes}
\item 
Note: The results are obtained from 1,000 simulations. See texts for more details.
\end{tablenotes}
\end{threeparttable}
\end{table}

\begin{table}
\setlength\extrarowheight{3pt}
\begin{threeparttable}
\centering
\caption{Simulation results under the Gumbel copula.\label{table:simu3}}
\begin{tabular*}{\textwidth}{@{\extracolsep{\fill}}ccrrrrrr}
\toprule
Kendall's Tau	&	URMC	&	\multicolumn{3}{c}{RISB}	&	\multicolumn{3}{c}{RIMSE}	\\
\cmidrule(lr){3-5}	\cmidrule(lr){6-8}
(Copula Parameter)	&	Estimator	&	\multicolumn{1}{c}{100}	&	\multicolumn{1}{c}{200}	&	\multicolumn{1}{c}{400}	&	\multicolumn{1}{c}{100}	&	\multicolumn{1}{c}{200}	&	\multicolumn{1}{c}{400}	\\
\midrule
	                    &	$\widehat u^\text{EBC}_{m^*,n}$         &	6.306	&	5.754	&	5.230	&	6.488	&	5.914	&	5.359	\\
	                    &	$\widehat u^\text{EBC}_{n^{1/2},n}$	    &	2.857	&	2.112	&	1.688	&	4.303	&	3.614	&	2.960	\\
$\tau_K=\frac{1}{3}$	&	$\widehat u^\beta$	                    &	0.562	&	0.255	&	0.253	&	9.975	&	8.836	&	7.875	\\
($\theta=1.5$)	        &	$\widehat u^\text{DR}_\text{Probit}$    &	2.414	&	2.236	&	2.192	&	4.626	&	3.612	&	3.031	\\
	                    &	$\widehat u^\text{DR}_\text{Logit}$   	&	2.349	&	2.135	&	2.070	&	4.472	&	3.450	&	2.875	\\
	                    &	$\widehat u^\text{DR}_\text{Logit,2}$	&	0.751	&	0.379	&	0.315	&	6.817	&	4.123	&	2.777	\\
\midrule
	                    &	$\widehat u^\text{EBC}_{m^*,n}$         &	7.885	&	7.048	&	6.262	&	8.091	&	7.219	&	6.408	\\
	                    &	$\widehat u^\text{EBC}_{n^{1/2},n}$	    &	4.333	&	3.215   &	2.560	&	5.169	&	4.149	&	3.383	\\
$\tau_K=\frac{1}{2}$	&	$\widehat u^\beta$	                    &	0.654	&	0.324	&	0.331	&	9.897	&	8.889	&	8.008	\\
($\theta=2$)	        &	$\widehat u^\text{DR}_\text{Probit}$    &	3.192	&	2.736	&	2.568	&	5.593	&	4.296	&	3.533	\\
	                    &	$\widehat u^\text{DR}_\text{Logit}$    	&	3.180	&	2.620	&	2.399	&	5.557	&	4.194	&	3.395	\\
	                    &	$\widehat u^\text{DR}_\text{Logit,2}$	&	1.289	&	0.656	&	0.252	&	8.159	&	4.991	&	3.380	\\
\midrule
	                    &	$\widehat u^\text{EBC}_{m^*,n}$         &	8.470	&	7.455	&	6.570	&	8.666	&	7.617	&	6.712	\\
	                    &	$\widehat u^\text{EBC}_{n^{1/2},n}$	    &	6.042	&	4.543	&	3.635	&	6.393	&	4.954	&	4.029	\\
$\tau_K=\frac{2}{3}$	&	$\widehat u^\beta$	                    &	0.975	&	0.531	&	0.344	&	8.984	&	8.474	&	7.739	\\
($\theta=3$)	        &	$\widehat u^\text{DR}_\text{Probit}$	&	4.434	&	3.180	&	2.610	&	7.442	&	5.333	&	4.015	\\
	                    &	$\widehat u^\text{DR}_\text{Logit}$	    &	4.635	&	3.229	&	2.535	&	7.625	&	5.434	&	4.036	\\
	                    &	$\widehat u^\text{DR}_\text{Logit,2}$	&	2.891	&	1.737	&	0.775	&	9.945	&	6.489	&	4.275	\\
\bottomrule
\end{tabular*}
\begin{tablenotes}
\item 
Note: The results are obtained from 1,000 simulations. See texts for more details.
\end{tablenotes}
\end{threeparttable}
\end{table}

\begin{table}
\setlength\extrarowheight{3pt}
\begin{threeparttable}
\centering
\caption{Simulation results under the independence copula.\label{table:simu4}}
\begin{tabular*}{\textwidth}{@{\extracolsep{\fill}}ccrrrrrr}
\toprule
Kendall's Tau	&	URMC	&	\multicolumn{3}{c}{RISB}	&	\multicolumn{3}{c}{RIMSE}	\\
\cmidrule(lr){3-5}	\cmidrule(lr){6-8}
\phantom{(Copula Parameter)}	&	Estimator	&	\multicolumn{1}{c}{100}	&	\multicolumn{1}{c}{200}	&	\multicolumn{1}{c}{400}	&	\multicolumn{1}{c}{100}	&	\multicolumn{1}{c}{200}	&	\multicolumn{1}{c}{400}	\\
\midrule
        	&	$\widehat u^\text{EBC}_{m^*,n}$         &	0.001	&	0.012	&	0.011	&	0.684	&	0.484	&	0.339	\\
        	&	$\widehat u^\text{EBC}_{n^{1/2},n}$	    &   0.072	&   0.041	&   0.069	&   3.365	&   2.940	&   2.439	\\
$\tau_K=0$	&	$\widehat u^\beta$	                    &	0.269	&	0.219	&	0.252	&	8.995	&	7.917	&	6.917	\\
	        &	$\widehat u^\text{DR}_\text{Probit}$	    &	0.415	&	0.214	&	0.101	&	2.872	&	1.978	&	1.369	\\
  	        &	$\widehat u^\text{DR}_\text{Logit}$	&	0.411	&	0.225	&	0.110	&	2.824	&	1.966	&	1.365	\\
        	&	$\widehat u^\text{DR}_\text{Logit,2}$	&	0.324	&	0.150	&	0.118	&	4.127	&	2.720	&	1.867	\\
\bottomrule
\end{tabular*}
\begin{tablenotes}
\item 
Note: The results are obtained from 1,000 simulations. See texts for more details.
\end{tablenotes}
\end{threeparttable}
\end{table}

\cref{table:simu1,table:simu2,table:simu3,table:simu4} report the simulation results for the root integrated squared bias (RISB) and root integrated mean squared error (RIMSE), defined as
\begin{align*}
\text{RISB}&=\sqrt{\int(\E(\widehat u(\tau,s))-u(\tau,s))^2\d s},\\
\text{RIMSE}&=\sqrt{\int\E((\widehat u(\tau,s)-u(\tau,s))^2)\d s},
\end{align*}
where $\widehat u(\tau,s)$ denotes a generic estimator. For numerical integration, we set $\tau=0$ and evaluate $s$ on the grid $\{0.01,0.02,\dotsc,0.99\}$. Each design is replicated 1,000 times for sample sizes $n\in\{100,200,400\}$.

From \cref{table:simu1,table:simu2,table:simu3}, it is unsurprising that the empirical beta copula-based estimator $\widehat u^\beta$ achieves the best RISB performance in most scenarios, as it exploits the maximal order $n$. Under the Gaussian copula, however, the DR estimator with a probit link, $\widehat u^\text{DR}_\text{Probit}$, can outperform $\widehat u^\beta$ particularly in large samples. By contrast, $\widehat u^\beta$ performs poorly in terms of RIMSE due to undersmoothing. The intermediate-order estimator $\widehat u^\text{EBC}_{n^{1/2},n}$ instead delivers consistently good performance, aside from the infeasible benchmark $\widehat u^\text{EBC}_{m^*,n}$. The DR estimators perform similarly to $\widehat u^\text{EBC}_{n^{1/2},n}$, with their RIMSE roughly halving as the sample size quadruples, suggesting convergence close to the parametric $\sqrt{n}$ rate despite misspecification of the conditional distribution. From \cref{table:simu4}, we further find that the DR estimators dominate both $\widehat u^\beta$ and $\widehat u^\text{EBC}_{n^{1/2},n}$ under the independence copula. Overall, these results indicate that DR estimators perform well when the link function is appropriately specified and the sample size is sufficiently large.

\begin{figure}
\centering
\includegraphics[width=0.5\textwidth]{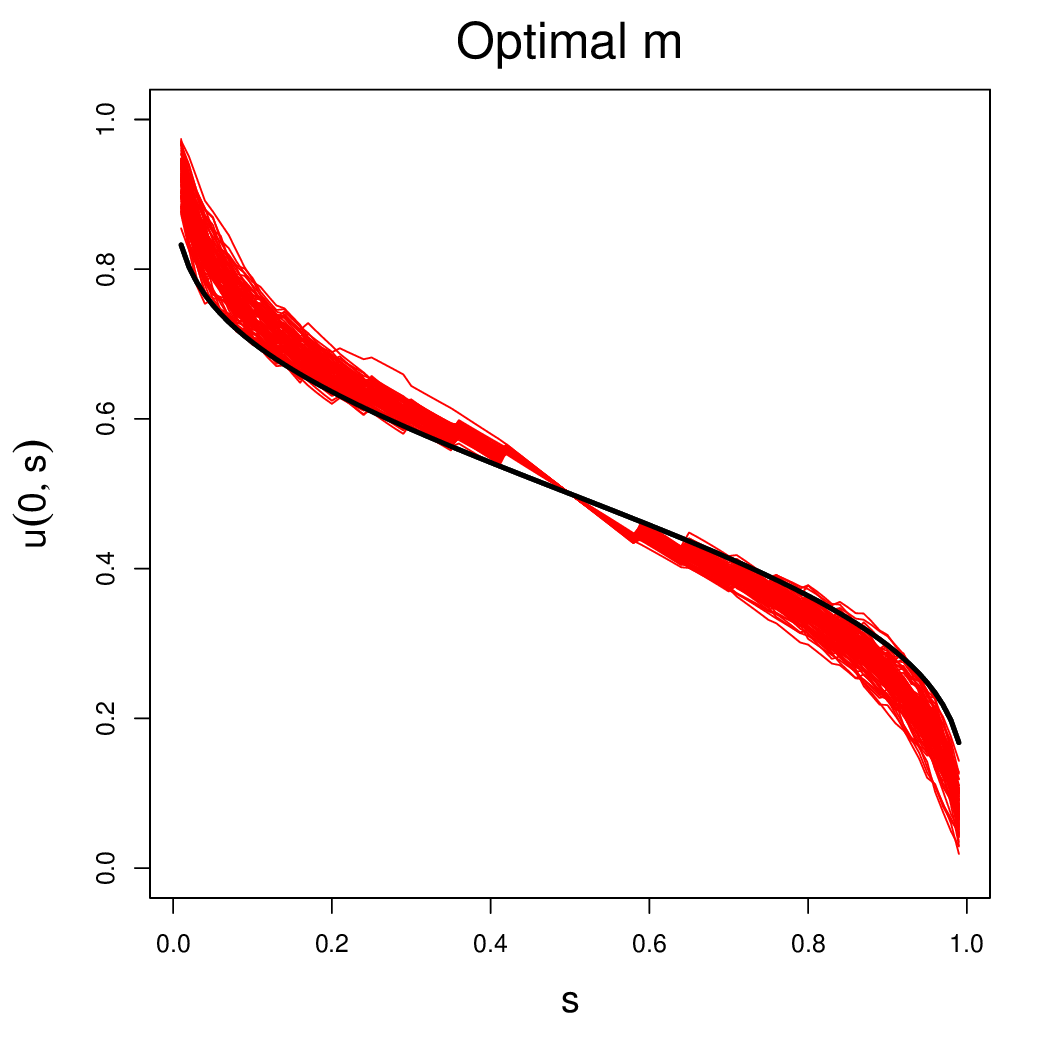}%
\includegraphics[width=0.5\textwidth]{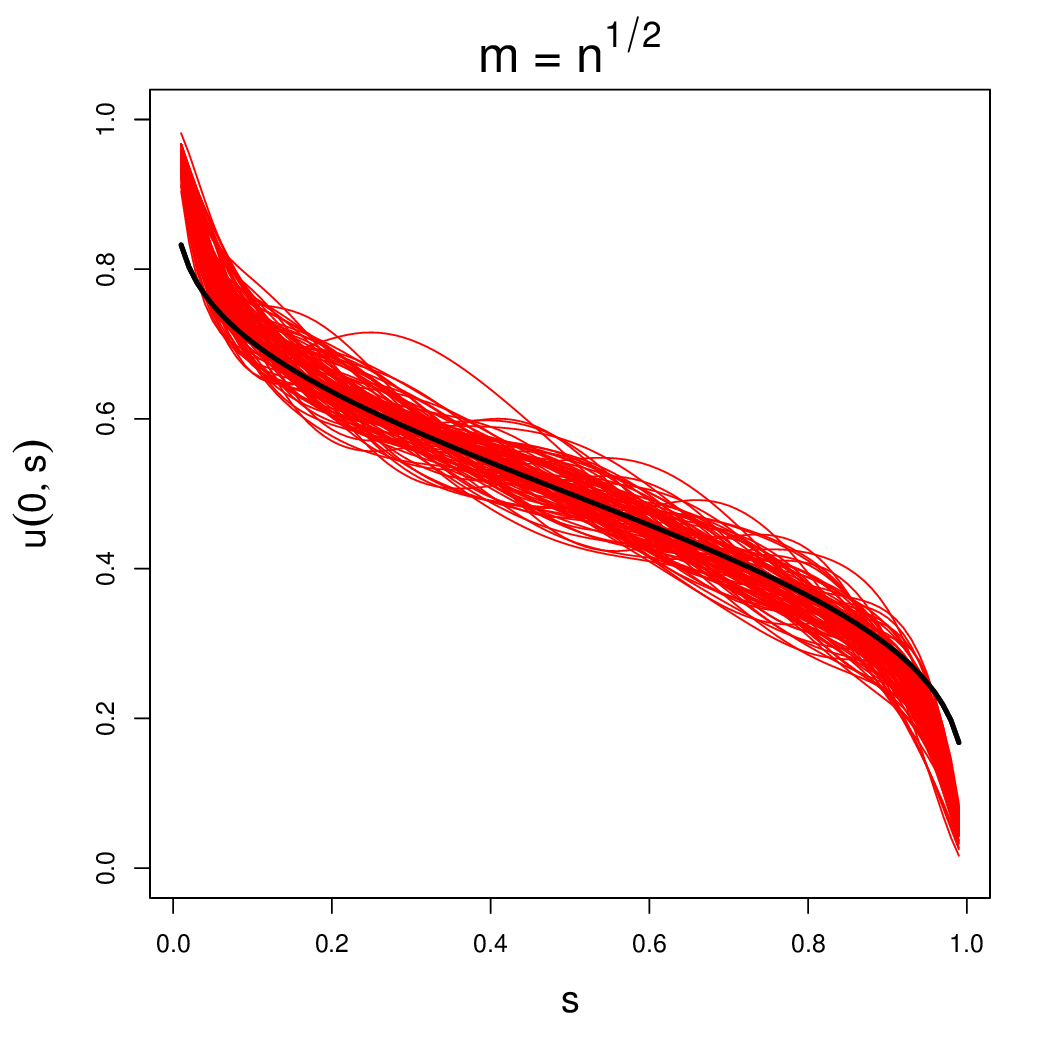}
\caption{Simulation results for estimating the upward rank mobility curve $u$ by the empirical Bernstein copula-based estimators $\widehat u^\text{EBC}_{m,n}$ using the optimal and $n^{1/2}$ orders under the Gaussian copula with 1,000 replications. The sample size is $n = 200$ and $\tau_K = 1/2$.\label{fig:gau}}
\end{figure}

\begin{figure}
\centering
\includegraphics[width=0.5\textwidth]{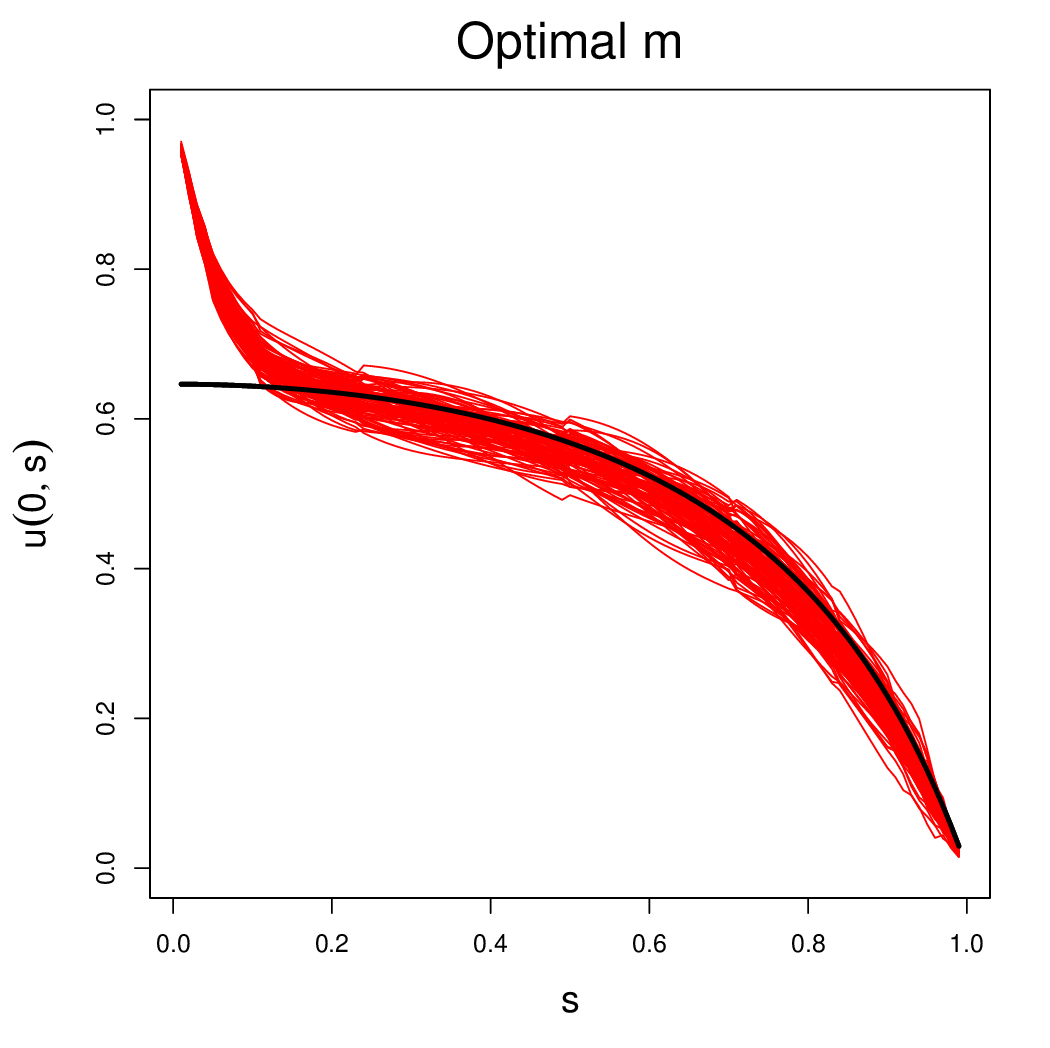}%
\includegraphics[width=0.5\textwidth]{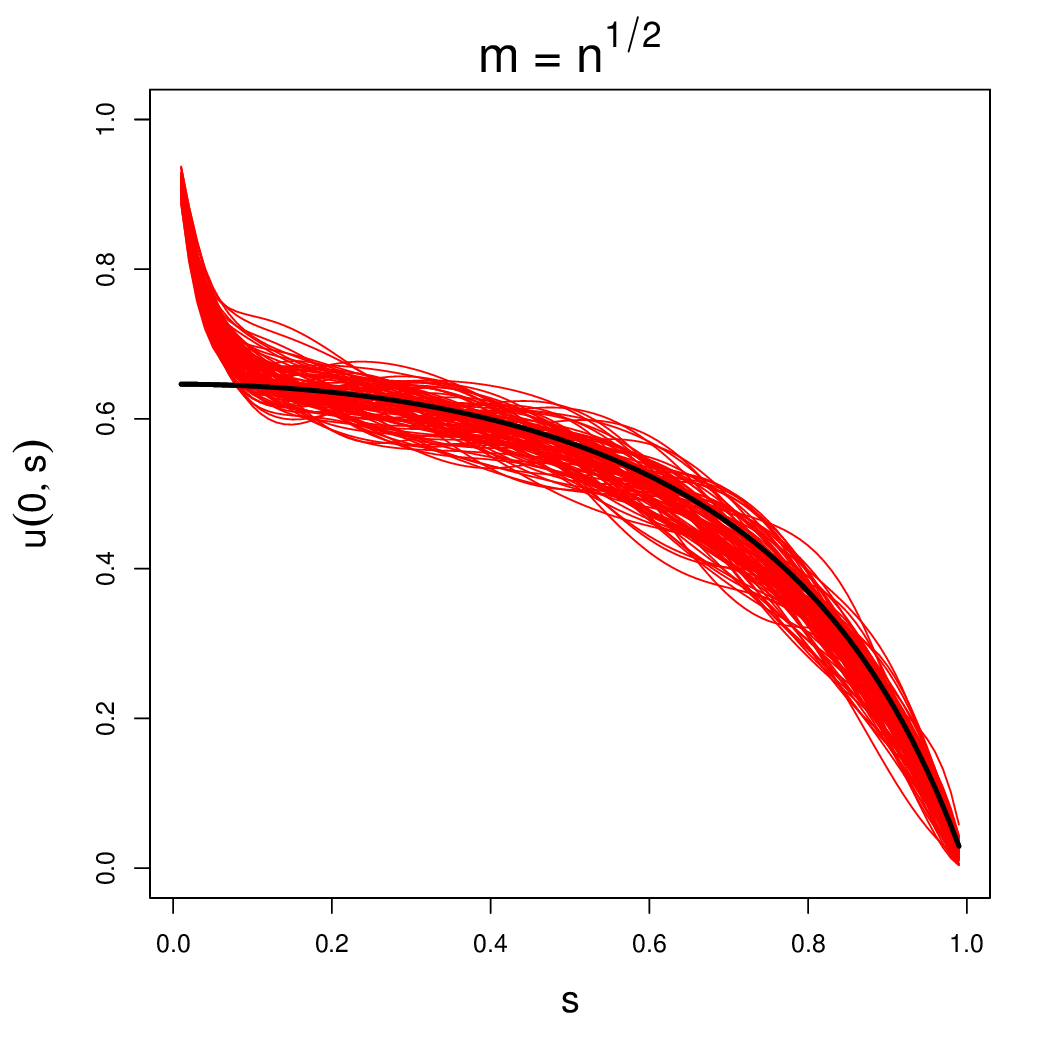}
\caption{Simulation results for estimating the upward rank mobility curve $u$ by the empirical Bernstein copula-based estimators $\widehat u^\text{EBC}_{m,n}$ using the optimal and $n^{1/2}$ orders under the Clayton copula with 1,000 replications. The sample size is $n = 200$ and $\tau_K = 1/2$.\label{fig:clay}}
\end{figure}

\begin{figure}
\centering
\includegraphics[width=0.5\textwidth]{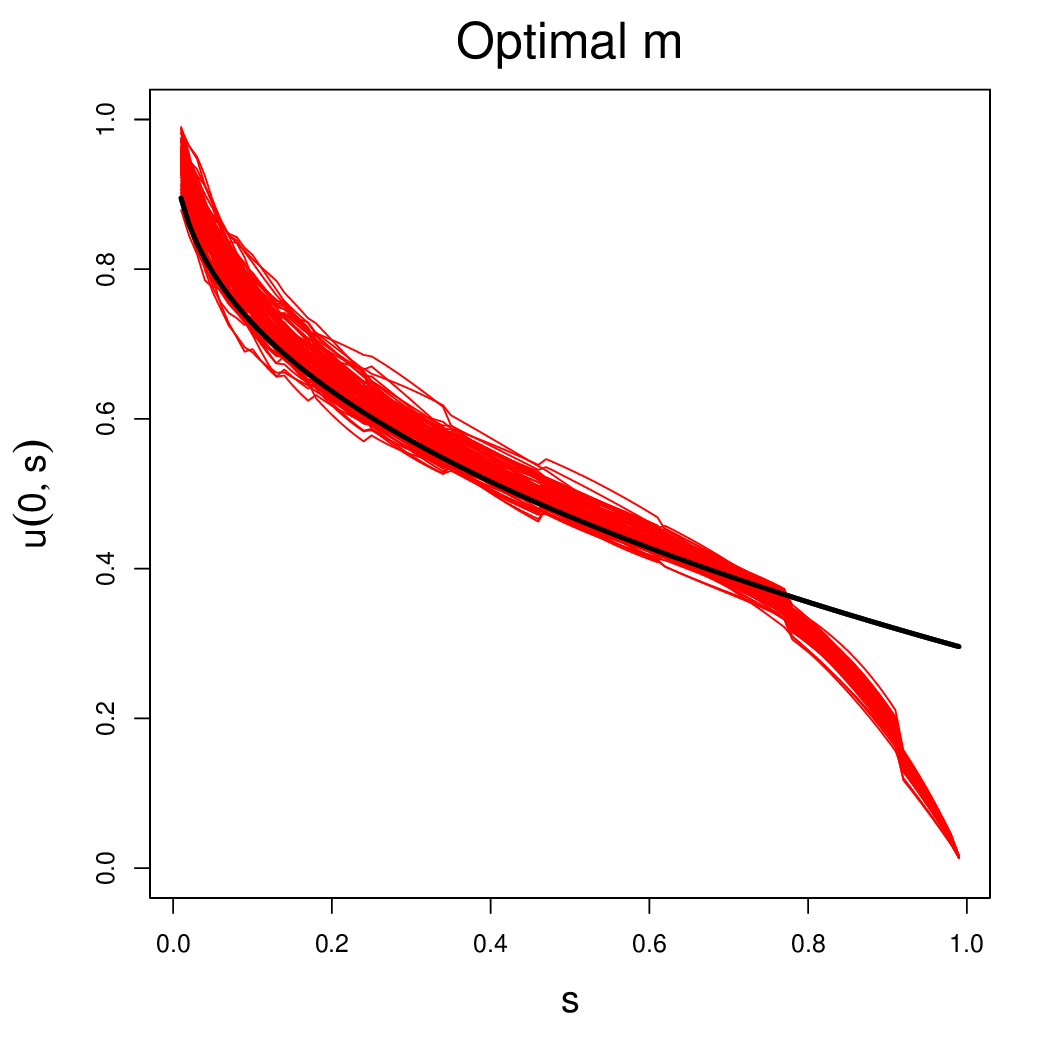}%
\includegraphics[width=0.5\textwidth]{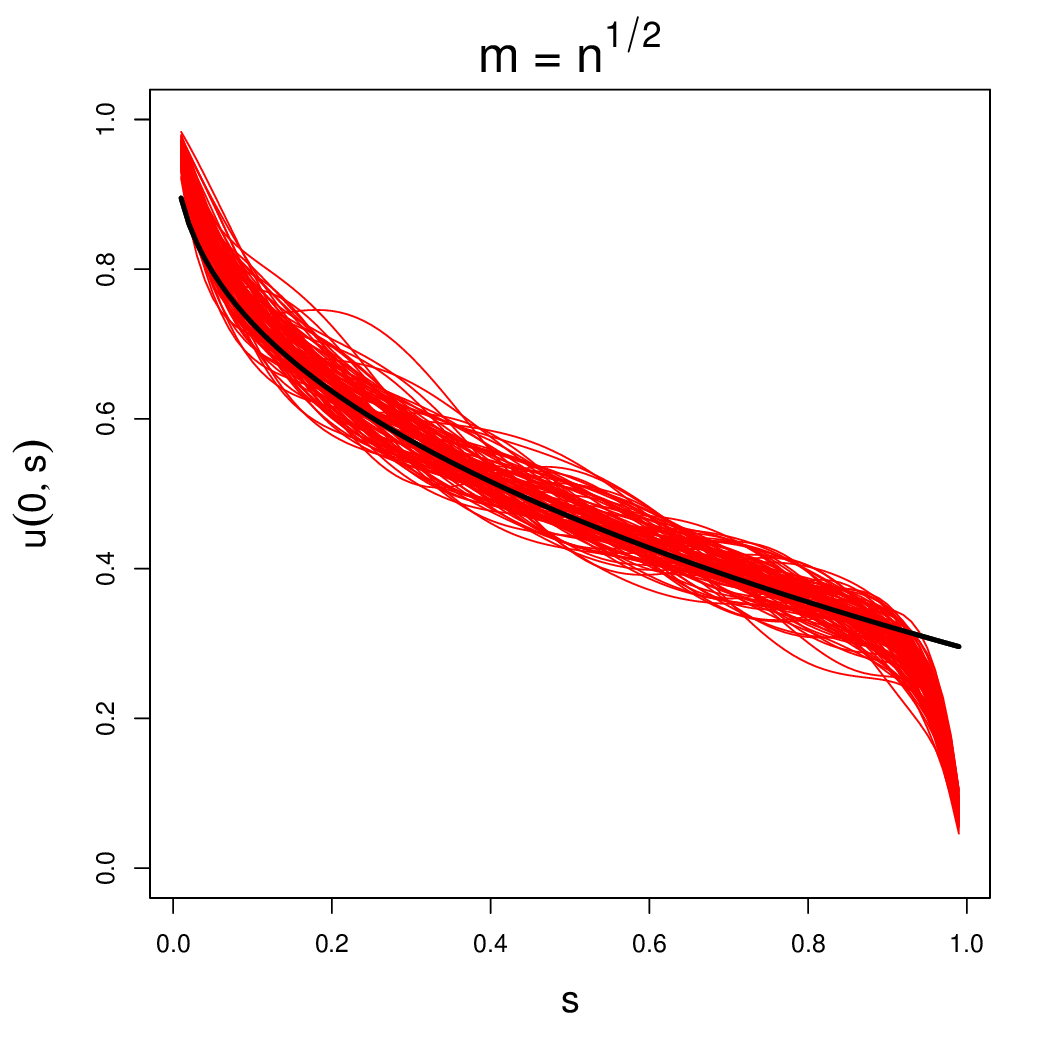}
\caption{Simulation results for estimating the upward rank mobility curve $u$ by the empirical Bernstein copula-based estimators $\widehat u^\text{EBC}_{m,n}$ using the optimal and $n^{1/2}$ orders under the Gumbel copula with 1,000 replications. The sample size is $n = 200$ and $\tau_K = 1/2$.\label{fig:gum}}
\end{figure}

Interestingly, the RIMSE of $\widehat u^\text{EBC}_{n^{1/2},n}$ can be smaller than that of $\widehat u^\text{EBC}_{m^,n}$ under both the Clayton and Gumbel copulas. To investigate this, we plot the estimated curves from simulation runs under the Gaussian, Clayton, and Gumbel copulas with sample size $n=200$ and Kendall's tau $\tau_K=1/2$. Under the Gaussian copula (\cref{fig:gau}), $\widehat u^\text{EBC}_{m^*,n}$ outperforms $\widehat u^\text{EBC}_{n^{1/2},n}$ because the bias function $b(s,s)$ defined in \cref{sec:ebc} is nearly zero around $s\approx 0.5$, with $b(0.5,0.5)=0$ exactly. Consequently, the optimal order $m^*(0,s)$ is small near $s\approx 0.5$, yielding very low variance and superior RIMSE of $\widehat u^\text{EBC}_{m^*,n}$. In contrast, $\widehat u^\text{EBC}_{n^{1/2},n}$ outperforms $\widehat u^\text{EBC}_{m^*,n}$ under the Clayton copula (\cref{fig:clay}) and the Gumbel copula (\cref{fig:gum}), as $\widehat u^\text{EBC}_{m^*,n}$ suffers from larger bias near $s\approx 0$ in the former and near $s\approx 1$ in the latter.

Taken together, these findings underscore that the optimal order $m^*$ depends on the shape of the curve and is derived from asymptotic mean squared error, which does not necessarily guarantee finite-sample optimality. Our simulations suggest that setting $m=n^{1/2}$ offers a practical and robust choice for the EBC estimator.

\section{Empirical Study}
\label{sec:empi}

In this section, we examine intergenerational upward rank mobility between whites and blacks in the United States. Following \citet{Bhattacharya2011}, we use data from the National Longitudinal Survey of Youth 1979 (NLSY79), which tracks individuals who were 14--22 years old in 1979 through adulthood until 2018. The survey provides information on annual income from the previous year. To avoid complications arising from labor force participation, we restrict the sample to 2,002 white sons and 276 black sons.

The parental permanent income is measured using total family income reported by the sons while living with their parents from 1979 to 1981, corresponding to income years 1978--1980. We subtract any recorded earnings of the sons (wages, salaries, farm or business income) and average the remaining values across all available years. The sons' permanent income is measured as the average of their annual earnings in 1998, 2000, 2002, and 2004, when they were approximately 40 years old. All income figures are adjusted to 1978 dollars using the CPI-U prior to averaging.\footnote{These settings are aligned with those in \citet{Bhattacharya2011}.}

\begin{figure}
\centering
\includegraphics[width=0.5\textwidth]{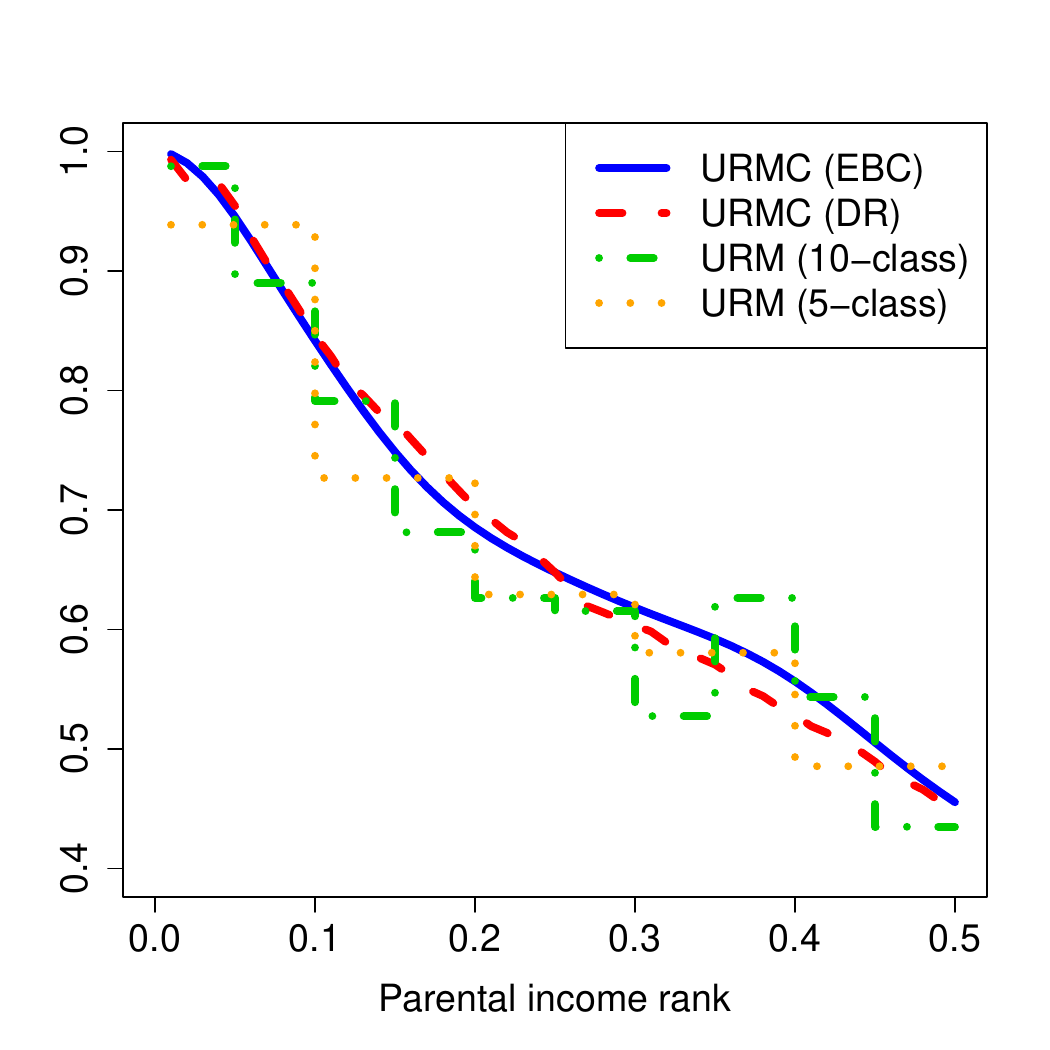}
\caption{URMCs and URMs.\label{fig:urmc-urm}}
\end{figure}

\cref{fig:urmc-urm} presents the estimated upward rank mobility curves (URMCs) obtained from the EBC and DR methods introduced in \cref{sec:ebc,sec:dr}. The blue solid line corresponds to the EBC estimator with $m=n^{1/2}$, as recommended by the simulation results in \cref{sec:simu}, while the red dashed line corresponds to the DR estimator with a logit link and quadratic polynomial specification. For reference, we also plot the sample analogues of upward rank mobility (URM) defined in \cref{urm}, constructed with different numbers of income categories (green and yellow step functions). The figure shows that both URMC estimates are reliable, as they closely match the empirical URMs for parental income ranks in $[0,0.5]$. More importantly, the URMCs reveal a monotonic decline in upward mobility as parental income rank increases---a pattern that the discrete URMs are unable to capture.

\begin{figure}
\centering
\includegraphics[width=0.5\textwidth]{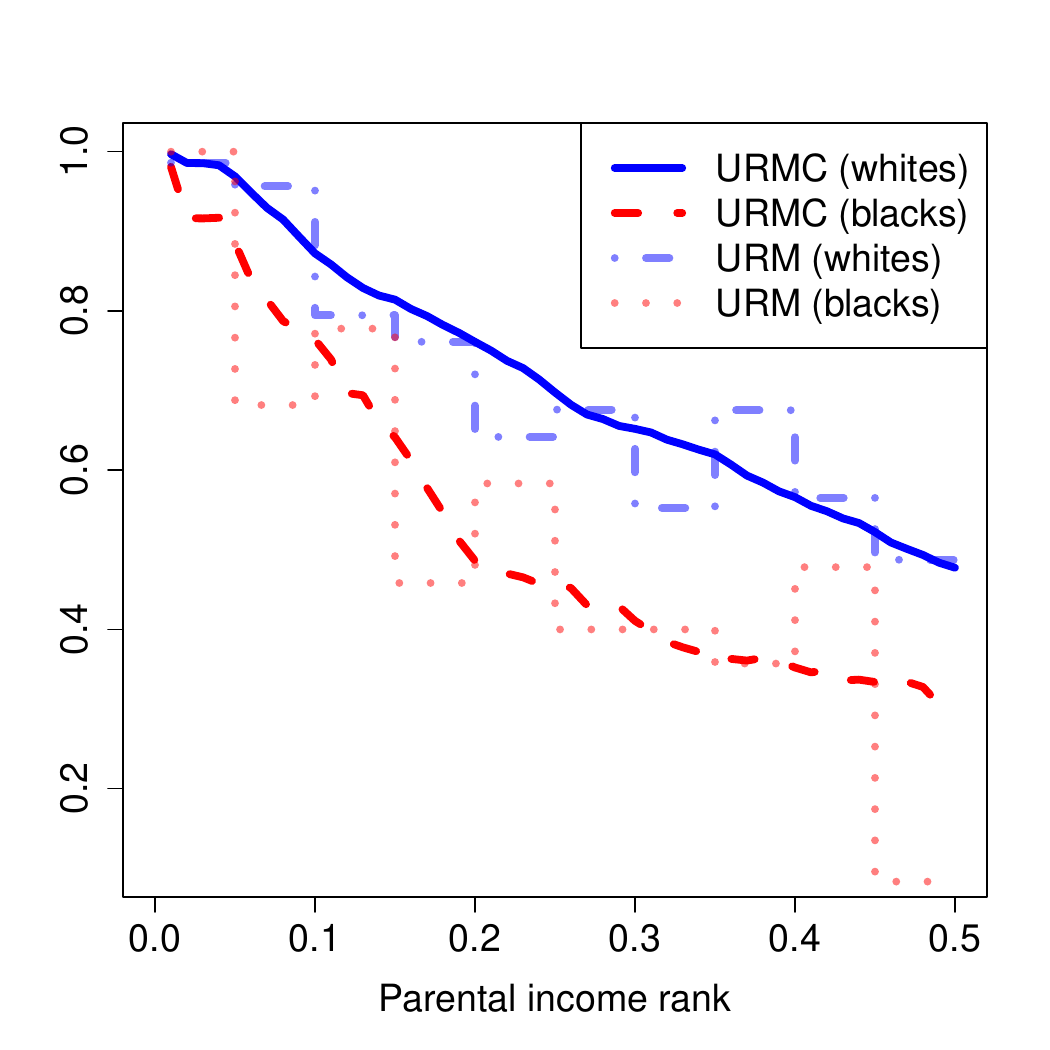}
\caption{URMCs and URMs by race.\label{fig:race}}
\end{figure}

\begin{figure}
\centering
\includegraphics[width=0.5\textwidth]{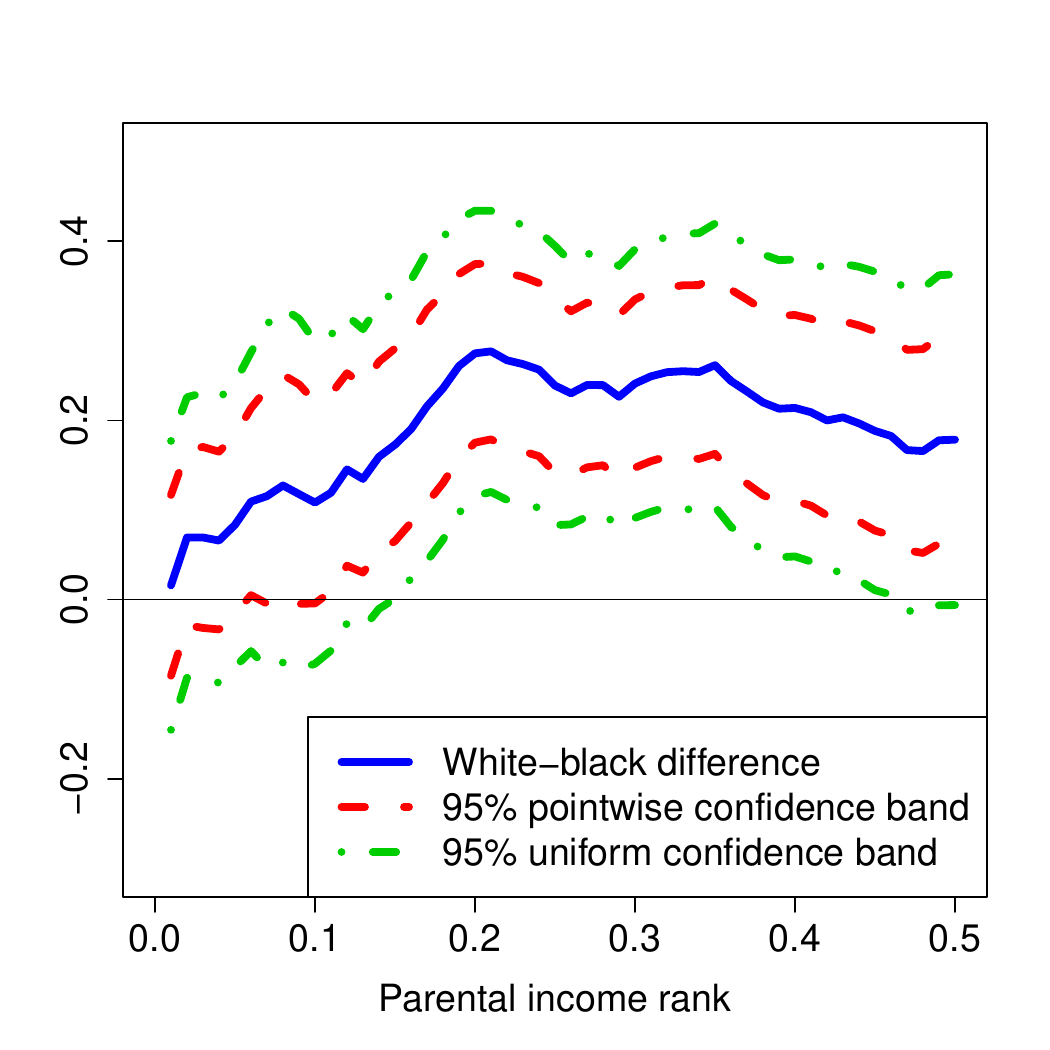}
\caption{White-black difference with 95\% pointwise and uniform confidence bands.\label{fig:diff}}
\end{figure}

We next examine differences in upward rank mobility between racial groups. Since the copula-based estimator is not applicable in this context, we employ the conditional URMC estimator from \cref{cond-dr} to construct race-specific mobility curves, as shown in \cref{fig:race}. The figure indicates that whites consistently exhibit higher upward rank mobility than blacks. Further evidence is provided in \cref{fig:diff}, which plots the white-black difference along with 95\% pointwise and uniform confidence bands over parental income ranks $[0,0.5]$. As the uniform band lies strictly above zero on the interval $[0.2,0.4]$, we conclude that whites statistically dominate blacks in upward rank mobility among lower-middle-income families in our sample.

\section{Conclusion}
\label{sec:conc}

This paper proposes the upward rank mobility curve as a tool for assessing intergenerational upward mobility across the entire parental income distribution. We show that the measure can be expressed solely in terms of the copula of parent and child income, and we develop nonparametric and semiparametric estimators for the measure and its conditional variant. An empirical application using the data of \citet{Bhattacharya2011} demonstrates systematically higher upward rank mobility for whites than for blacks in the United States. We further provide evidence of significant upward mobility dominance of whites over blacks among lower-middle-income families.

\clearpage

\bibliographystyle{ecta}
\bibliography{URMC_bib}

\clearpage

\appendix

\section{Proofs}
\label{sec:proof}

\subsection*{Proof of \cref{pro:urmc}}

\begin{proof}
Since $F_0$ and $F_1$ are continuous, Sklar's theorem \citep{Sklar1959} states that there exists a unique copula function $C$ coupling the joint CDF of $(Y_0,Y_1)$ to its univariate margins,
\[
\Prob(Y_0\leq y_0,Y_1\leq y_1)=C(F_0(y_0),F_1(y_1)).
\]
If, additionally, assume that $F_0$ and $F_1$ are strictly increasing, one can take the partial derivative with respect to $y_0$ to yield
\[
\Prob(Y_1\leq y_1|Y_0=y_0)f_0(y_0)=\partial_0C(F_0(y_0),F_1(y_1))f_0(y_0)\quad a.e.,
\]
where $f_0$ is the probability density function (PDF) of $Y_0$ and $\partial_0C=\partial C(u_0,u_1)/\partial u_0$ denotes the first partial derivative of $C$. Divide by $f_0$ on both sides and let $(y_0,y_1)=(Q_0(s),Q_1(s+\tau))$, where $Q_j=F^{-1}_j$ is the quantile function of $Y_j$ for $j=0,1$, we have
\[
\Prob(Y_1\leq Q_1(s+\tau)|Y_0=Q_0(s))=\partial_0C(s,s+\tau)\quad a.e.
\]
Thus, \cref{urmc} can be rewritten as
\begin{align*}
u(\tau,s)
&=\Prob(F_1(Y_1)>s+\tau|F_0(Y_0)=s)\\
&=\Prob(Y_1>Q_1(s+\tau)|Y_0=Q_0(s))\\
&=1-\Prob(Y_1\leq Q_1(s+\tau)|Y_0=Q_0(s))\\
&=1-\partial_0C(s,s+\tau)\quad a.e.\tag*{\qedhere}
\end{align*}
\end{proof}

\subsection*{Proof of \cref{thm:weak}}

First, \cref{urmc-dr} and simple algebra give
\begin{align}
&\sqrt{n}\left(\widehat u^\text{DR}(\tau,s)-u(\tau,s)\right)\notag\\
&=-\left\{\sqrt{n}\left(\widehat F_{1|0}(Q_1(s+\tau)|Q_0(s))-F_{1|0}(Q_1(s+\tau)|Q_0(s))\right)\right.\label{expan:F}\\
&\quad+\sqrt{n}\left(\widehat F_{1|0}(\widehat Q_1(s+\tau)|Q_0(s))-\widehat F_{1|0}(Q_1(s+\tau)|Q_0(s))\right)\label{expan:Q1}\\
&\quad\left.+\sqrt{n}\left(\widehat F_{1|0}(\widehat Q_1(s+\tau)|\widehat Q_0(s))-\widehat F_{1|0}(\widehat Q_1(s+\tau)|Q_0(s))\right)\right\}.\label{expan:Q0}
\end{align}
By Corollary 5.4 of \cite*{Chernozhukov2013}, the term inside the curly brackets in \cref{expan:F} is asymptotically equivalent to $n^{-1/2}\sum_{i=1}^n\ell(Q_1(s+\tau),Q_0(s),Y_{1i},Y_{0i})$, where
\begin{equation}
\ell(y_1,y_0,Y_1,Y_0)\equiv\lambda(P(y_0)^\top\theta(y_1))P(y_0)^\top
\psi_\theta(y_1,Y_1,Y_0),\label{infl:F}
\end{equation}
and $\{\ell(y_1,y_0,Y_1,Y_0):(y_1,y_0)\in\Y_1\times\Y_0\}$ belongs to some Donsker class. This Donsker property implies stochastic equicontinuity of $\sqrt{n}(\widehat F_{1|0}(y_1|y_0)-F_{1|0}(y_1|y_0))$. That is, for all $\eta>0$ and $\epsilon>0$, there exists $\delta>0$ such that
\begin{multline}
\limsup_{n\to\infty}\Prob\left(\sup_{|y_0-y_0'|+|y_1-y_1'|\leq\delta}\left|\sqrt{n}\left(\widehat F_{1|0}(y_1|y_0)-F_{1|0}(y_1|y_0)\right)-\right.\right.\\
\left.\left.\sqrt{n}\left(\widehat F_{1|0}(y_1'|y_0')-F_{1|0}(y_1'|y_0')\right)\right|>\eta\right)\leq\epsilon.
\end{multline}

Given this, \cref{expan:Q1} can be rewritten as
\begin{align}
&\sqrt{n}\left(\widehat F_{1|0}(\widehat Q_1(s+\tau)|Q_0(s))-\widehat F_{1|0}(Q_1(s+\tau)|Q_0(s))\right)\notag\\
&=\sqrt{n}\left(F_{1|0}(\widehat Q_1(s+\tau)|Q_0(s))-F_{1|0}(Q_1(s+\tau)|Q_0(s))\right)+o_p(1)\notag\\
&=\sqrt{n}\left(\Lambda(P(Q_0(s))^\top\theta(\widehat Q_1(s+\tau)))-\Lambda(P(Q_0(s))^\top\theta(Q_1(s+\tau)))\right)+o_p(1)\notag\\
&=\lambda(P(Q_0(s))^\top\theta(Q_1(s+\tau)))P(Q_0(s))^\top\theta'(Q_1(s+\tau))\cdot\notag\\
&\pushright{\sqrt{n}\left(\widehat Q_1(s+\tau)-Q_1(s+\tau)\right)+o_p(1)}\notag\\
&=\lambda(P(Q_0(s))^\top\theta(Q_1(s+\tau)))P(Q_0(s))^\top\theta'(Q_1(s+\tau))\cdot\notag\\
&\pushright{\left\{\frac{1}{\sqrt{n}}\sum_{i=1}^n\frac{(s+\tau)-\1(Y_1\leq Q_1(s+\tau))}{f_1(Q_1(s+\tau))}\right\}+o_p(1)}\notag\\
&\equiv\frac{1}{\sqrt{n}}\sum_{i=1}^n\lambda(P(Q_0(s))^\top\theta(Q_1(s+\tau)))P(Q_0(s))^\top\theta'(Q_1(s+\tau))\psi_{1}(\tau,s,Y_1)+o_p(1),\label{infl:Q1}
\end{align}
where the first equality holds by the stochastic equicontinuity, the second equality holds by the assumption that $F_{1|0}(y_1|y_0)=\Lambda(P(y_0)^\top\theta(y_1))$, and the third and fourth equalities hold by the functional delta method and $Q_1=F_1^{-1}$. Note that \cref{infl:Q1} also belongs to some Donsker class of functions under \cref{asm:dr}.

Finally, by similar arguments we can show that \cref{expan:Q0} is asymptotically equivalent to
\begin{equation}
\lambda(P(Q_0(s))^\top\theta(Q_1(s+\tau)))P'(Q_0(s))^\top\theta(Q_1(s+\tau))\psi_{0,s}(Y_0),\label{infl:Q0}
\end{equation}
which is Donsker as well under \cref{asm:dr}. \cref{thm:weak} then follows by the functional central limit theorem.\hfill\qed

\end{document}

%% file: preamble.tex
\usepackage{amssymb,amsthm,bm,bbm,microtype,xcolor,natbib}
\usepackage[fleqn]{mathtools}
\usepackage[shortlabels]{enumitem}
\usepackage[margin=1in]{geometry}
\usepackage[doublespacing]{setspace}
\usepackage[colorlinks,allcolors=blue!50!black]{hyperref}
\usepackage[capitalise,noabbrev]{cleveref}
\usepackage[title]{appendix}

\usepackage{booktabs,multirow,graphicx,epstopdf,dcolumn}
\newcolumntype{.}[1]{D{.}{.}{#1}}
\newcolumntype{d}[1]{D{.}{.}{#1}}
\usepackage[flushleft]{threeparttable}
\usepackage[labelformat=simple]{subcaption}

\usepackage[labelsep=period]{caption}

\makeatletter
\newcommand{\pushleft}[1]{\ifmeasuring@#1\else\omit$\displaystyle#1$\hfill\fi\ignorespaces}
\newcommand{\pushright}[1]{\ifmeasuring@#1\else\omit\hfill$\displaystyle#1$\fi\ignorespaces}
\makeatother

\let\originalleft\left
\let\originalright\right
\renewcommand{\left}{\mathopen{}\mathclose\bgroup\originalleft}
\renewcommand{\right}{\aftergroup\egroup\originalright}

\DeclareFontFamily{U}{mathx}{\hyphenchar\font45}
\DeclareFontShape{U}{mathx}{m}{n}{<-> mathx10}{}
\DeclareSymbolFont{mathx}{U}{mathx}{m}{n}
\DeclareMathAccent{\widebar}{0}{mathx}{"73}

\DeclareMathOperator{\E}{E}

\DeclareMathOperator{\Prob}{P}
\DeclareMathOperator{\AMSE}{AMSE}

\newcommand{\1}{1}

\newcommand{\Y}{\mathcal{Y}}
\newcommand{\R}{\mathbb{R}}

\renewcommand{\d}[1]{\ensuremath{\operatorname{d}\!{#1}}}

\theoremstyle{plain}
\newtheorem{asm}{Assumption}

\newtheorem{thm}{Theorem}

\newtheorem{pro}{Proposition}

\newtheorem{defn}{Definition}
\theoremstyle{definition}
\newtheorem{rmk}{Remark}

\newlist{asmenum}{enumerate}{1}
\setlist[asmenum]{label=(\roman*),ref=\theasm{}(\roman*)}
\crefalias{asmenumi}{asm}
\crefname{asm}{Assumption}{Assumptions}
\crefname{lem}{Lemma}{Lemmas}
\crefname{thm}{Theorem}{Theorems}
\crefname{cor}{Corollary}{Corollaries}
\crefname{pro}{Proposition}{Propositions}
\crefname{rmk}{Remark}{Remarks}
\crefname{equation}{}{}

%% file: URMC.bbl
\begin{thebibliography}{39}
\newcommand{\enquote}[1]{``#1''}
\expandafter\ifx\csname natexlab\endcsname\relax\def\natexlab#1{#1}\fi

\bibitem[\protect\citeauthoryear{Aaberge and Mogstad}{Aaberge and
  Mogstad}{2014}]{Aaberge2014}
\textsc{Aaberge, R. and M.~Mogstad} (2014): \enquote{Income mobility as an
  equalizer of permanent income,} Tech. rep.

\bibitem[\protect\citeauthoryear{Bavaro and Tullio}{Bavaro and
  Tullio}{2023}]{Bavaro2023}
\textsc{Bavaro, M. and F.~Tullio} (2023): \enquote{Intergenerational mobility
  measurement with latent transition matrices,} \emph{The Journal of Economic
  Inequality}, 21, 25--45.

\bibitem[\protect\citeauthoryear{Bhattacharya and Mazumder}{Bhattacharya and
  Mazumder}{2011}]{Bhattacharya2011}
\textsc{Bhattacharya, D. and B.~Mazumder} (2011): \enquote{A nonparametric
  analysis of black-white differences in intergenerational income mobility in
  the United States,} \emph{Quantitative Economics}, 2, 335--379.

\bibitem[\protect\citeauthoryear{Black and Devereux}{Black and
  Devereux}{2011}]{Black2011}
\textsc{Black, S.~E. and P.~J. Devereux} (2011): \emph{Recent Developments in
  Intergenerational Mobility}, Elsevier, vol.~4, 1487--1541.

\bibitem[\protect\citeauthoryear{Bratberg, Davis, Mazumder, Nybom, Schnitzlein,
  and Vaage}{Bratberg et~al.}{2017}]{Bratberg2017}
\textsc{Bratberg, E., J.~Davis, B.~Mazumder, M.~Nybom, D.~D. Schnitzlein, and
  K.~Vaage} (2017): \enquote{A Comparison of Intergenerational Mobility Curves
  in Germany, Norway, Sweden, and the US,} \emph{The Scandinavian Journal of
  Economics}, 119, 72--101.

\bibitem[\protect\citeauthoryear{Callaway, Li, and Murtazashvili}{Callaway
  et~al.}{2024}]{Callaway2024}
\textsc{Callaway, B., T.~Li, and I.~Murtazashvili} (2024):
  \enquote{Distributional Effects with Two-Sided Measurement Error: An
  Application to Intergenerational Income Mobility *,} Tech. rep.

\bibitem[\protect\citeauthoryear{Chernozhukov, Fernández-Val, Meier, van
  Vuuren, and Vella}{Chernozhukov et~al.}{2025}]{Chernozhukov2025}
\textsc{Chernozhukov, V., I.~Fernández-Val, J.~Meier, A.~van Vuuren, and
  F.~Vella} (2025): \enquote{Bivariate Distribution Regression; Theory,
  Estimation and an Application to Intergenerational Mobility,} .

\bibitem[\protect\citeauthoryear{Chernozhukov, Fernández-Val, and
  Melly}{Chernozhukov et~al.}{2013}]{Chernozhukov2013}
\textsc{Chernozhukov, V., I.~Fernández-Val, and B.~Melly} (2013):
  \enquote{Inference on Counterfactual Distributions,} \emph{Econometrica}, 81,
  2205--2268.

\bibitem[\protect\citeauthoryear{Chetty, Grusky, Hell, Hendren, Manduca, and
  Narang}{Chetty et~al.}{2017}]{Chetty2017}
\textsc{Chetty, R., D.~Grusky, M.~Hell, N.~Hendren, R.~Manduca, and J.~Narang}
  (2017): \enquote{The fading American dream: Trends in absolute income
  mobility since 1940,} \emph{Science}, 356, 398--406.

\bibitem[\protect\citeauthoryear{Chetty, Hendren, Jones, and Porter}{Chetty
  et~al.}{2019}]{Chetty2019}
\textsc{Chetty, R., N.~Hendren, M.~R. Jones, and S.~R. Porter} (2019):
  \enquote{Race and Economic Opportunity in the United States: an
  Intergenerational Perspective*,} \emph{The Quarterly Journal of Economics},
  135, 711--783.

\bibitem[\protect\citeauthoryear{Chetty, Hendren, Kline, and Saez}{Chetty
  et~al.}{2014}]{Chetty2014}
\textsc{Chetty, R., N.~Hendren, P.~Kline, and E.~Saez} (2014): \enquote{Where
  is the land of Opportunity? The Geography of Intergenerational Mobility in
  the United States *,} \emph{The Quarterly Journal of Economics}, 129,
  1553--1623.

\bibitem[\protect\citeauthoryear{Collins and Wanamaker}{Collins and
  Wanamaker}{2022}]{Collins2022}
\textsc{Collins, W.~J. and M.~H. Wanamaker} (2022): \enquote{African American
  Intergenerational Economic Mobility since 1880,} \emph{American Economic
  Journal: Applied Economics}, 14, 84--117.

\bibitem[\protect\citeauthoryear{Corak}{Corak}{2013}]{Corak2013}
\textsc{Corak, M.} (2013): \enquote{Income Inequality, Equality of Opportunity,
  and Intergenerational Mobility,} \emph{Journal of Economic Perspectives}, 27,
  79--102.

\bibitem[\protect\citeauthoryear{Corak}{Corak}{2020}]{Corak2020}
---\hspace{-.1pt}---\hspace{-.1pt}--- (2020): \enquote{The Canadian Geography
  of Intergenerational Income Mobility,} \emph{The Economic Journal}, 130,
  2134--2174.

\bibitem[\protect\citeauthoryear{Corak, Lindquist, and Mazumder}{Corak
  et~al.}{2014}]{Corak2014}
\textsc{Corak, M., M.~J. Lindquist, and B.~Mazumder} (2014): \enquote{A
  comparison of upward and downward intergenerational mobility in Canada,
  Sweden and the United States,} \emph{Labour Economics}, 30, 185--200.

\bibitem[\protect\citeauthoryear{Dardanoni, Fiorini, and Forcina}{Dardanoni
  et~al.}{2012}]{Dardanoni2012}
\textsc{Dardanoni, V., M.~Fiorini, and A.~Forcina} (2012): \enquote{Stochastic
  monotonicity in intergenerational mobility tables,} \emph{Journal of Applied
  Econometrics}, 27, 85--107.

\bibitem[\protect\citeauthoryear{Davies and Shorrocks}{Davies and
  Shorrocks}{1989}]{Davies1989}
\textsc{Davies, J.~B. and A.~F. Shorrocks} (1989): \enquote{Optimal grouping of
  income and wealth data,} \emph{Journal of Econometrics}, 42, 97--108.

\bibitem[\protect\citeauthoryear{Deutscher and Mazumder}{Deutscher and
  Mazumder}{2023}]{Deutscher2023}
\textsc{Deutscher, N. and B.~Mazumder} (2023): \enquote{Measuring
  Intergenerational Income Mobility: A Synthesis of Approaches,} \emph{Journal
  of Economic Literature}, 61, 988--1036.

\bibitem[\protect\citeauthoryear{Durlauf}{Durlauf}{2004}]{Durlauf2004}
\textsc{Durlauf, S.~N.} (2004): \enquote{Neighborhood Effects,} \emph{Handbook
  of Regional and Urban Economics}, 4, 2173--2242.

\bibitem[\protect\citeauthoryear{Fermanian, Radulovic, and Wegkamp}{Fermanian
  et~al.}{2004}]{Fermanian2004}
\textsc{Fermanian, J.-D., D.~Radulovic, and M.~Wegkamp} (2004): \enquote{Weak
  convergence of empirical copula processes,} \emph{Bernoulli}, 10.

\bibitem[\protect\citeauthoryear{Fields, Leary, and Ok}{Fields
  et~al.}{2002}]{Fields2002}
\textsc{Fields, G.~S., J.~B. Leary, and E.~A. Ok} (2002): \enquote{Stochastic
  dominance in mobility analysis,} \emph{Economics Letters}, 75, 333--339.

\bibitem[\protect\citeauthoryear{Foresi and Peracchi}{Foresi and
  Peracchi}{1995}]{Foresi1995}
\textsc{Foresi, S. and F.~Peracchi} (1995): \enquote{The Conditional
  Distribution of Excess Returns: An Empirical Analysis,} \emph{Journal of the
  American Statistical Association}, 90, 451--466.

\bibitem[\protect\citeauthoryear{Formby, Smith, and Zheng}{Formby
  et~al.}{2004}]{Formby2004}
\textsc{Formby, J.~P., W.~Smith, and B.~Zheng} (2004): \enquote{Mobility
  measurement, transition matrices and statistical inference,} \emph{Journal of
  Econometrics}, 120, 181--205.

\bibitem[\protect\citeauthoryear{Fox}{Fox}{2016}]{Fox2016}
\textsc{Fox, L.~E.} (2016): \enquote{Parental Wealth and the Black–White
  Mobility Gap in the U.S.} \emph{Review of Income and Wealth}, 62, 706--723.

\bibitem[\protect\citeauthoryear{Grawe}{Grawe}{2004}]{Grawe2004}
\textsc{Grawe, N.~D.} (2004): \enquote{Reconsidering the Use of Nonlinearities
  in Intergenerational Earnings Mobility as a Test for Credit Constraints,}
  Tech. rep.

\bibitem[\protect\citeauthoryear{Hnatkovska, Lahiri, and Paul}{Hnatkovska
  et~al.}{2013}]{Hnatkovska2013}
\textsc{Hnatkovska, V., A.~Lahiri, and S.~B. Paul} (2013): \enquote{Breaking
  the Caste Barrier: Intergenerational Mobility in India,} \emph{Journal of
  Human Resources}, 48, 435--473.

\bibitem[\protect\citeauthoryear{Janssen, Swanepoel, and Veraverbeke}{Janssen
  et~al.}{2016}]{Janssen2016}
\textsc{Janssen, P., J.~Swanepoel, and N.~Veraverbeke} (2016):
  \enquote{Bernstein estimation for a copula derivative with application to
  conditional distribution and regression functionals,} \emph{TEST}, 25,
  351--374.

\bibitem[\protect\citeauthoryear{Jappelli and Pistaferri}{Jappelli and
  Pistaferri}{2006}]{Jappelli2006}
\textsc{Jappelli, T. and L.~Pistaferri} (2006): \enquote{Intertemporal Choice
  and Consumption Mobility,} \emph{Journal of the European Economic
  Association}, 4, 75--115.

\bibitem[\protect\citeauthoryear{Lee, Linton, and Whang}{Lee
  et~al.}{2009}]{Lee2009}
\textsc{Lee, S., O.~Linton, and Y.~J. Whang} (2009): \enquote{Testing for
  Stochastic Monotonicity,} \emph{Econometrica}, 77, 585--602.

\bibitem[\protect\citeauthoryear{Lerman and Yitzhaki}{Lerman and
  Yitzhaki}{1989}]{Lerman1989}
\textsc{Lerman, R.~I. and S.~Yitzhaki} (1989): \enquote{Improving the accuracy
  of estimates of Gini coefficients,} \emph{Journal of Econometrics}, 42,
  43--47.

\bibitem[\protect\citeauthoryear{Mazumder}{Mazumder}{2014}]{Mazumder2014}
\textsc{Mazumder, B.} (2014): \enquote{Black–white differences in
  intergenerational economic mobility in the United States,} \emph{Economic
  Perspectives}, 38.

\bibitem[\protect\citeauthoryear{Millimet, Li, and Roychowdhury}{Millimet
  et~al.}{2020}]{Millimet2020}
\textsc{Millimet, D.~L., H.~Li, and P.~Roychowdhury} (2020): \enquote{Partial
  Identification of Economic Mobility: With an Application to the United
  States,} \emph{Journal of Business \& Economic Statistics}, 38, 732--753,
  doi: 10.1080/07350015.2019.1569527.

\bibitem[\protect\citeauthoryear{Richey and Rosburg}{Richey and
  Rosburg}{2018}]{Richey2018}
\textsc{Richey, J. and A.~Rosburg} (2018): \enquote{Decomposing economic
  mobility transition matrices,} \emph{Journal of Applied Econometrics}, 33,
  91--108.

\bibitem[\protect\citeauthoryear{Sancetta and Satchell}{Sancetta and
  Satchell}{2004}]{Sancetta2004}
\textsc{Sancetta, A. and S.~Satchell} (2004): \enquote{The Bernstein copula and
  its applications to modeling and approximations of multivariate
  distributions,} \emph{Econometric Theory}, 20, 535--562.

\bibitem[\protect\citeauthoryear{Segers, Sibuya, and Tsukahara}{Segers
  et~al.}{2017}]{Segers2017}
\textsc{Segers, J., M.~Sibuya, and H.~Tsukahara} (2017): \enquote{The empirical
  beta copula,} \emph{Journal of Multivariate Analysis}, 155, 35--51.

\bibitem[\protect\citeauthoryear{Sklar}{Sklar}{1959}]{Sklar1959}
\textsc{Sklar, A.} (1959): \enquote{Fonctions de répartition à n dimensions
  et leurs marges,} \emph{Publications de l' Institut Statistique de
  l'Université de Paris}, 8, 229--231.

\bibitem[\protect\citeauthoryear{Solon}{Solon}{1992}]{Solon1992}
\textsc{Solon, G.} (1992): \enquote{Intergenerational Income Mobility in the
  United States,} \emph{The American Economic Review}, 82, 393--408.

\bibitem[\protect\citeauthoryear{Solon}{Solon}{1999}]{Solon1999}
---\hspace{-.1pt}---\hspace{-.1pt}--- (1999): \emph{Intergenerational Mobility
  in the Labor Market}, Elsevier, vol.~3, 1761--1800.

\bibitem[\protect\citeauthoryear{Swanepoel and Allison}{Swanepoel and
  Allison}{2013}]{Swanepoel2013}
\textsc{Swanepoel, J. W.~H. and J.~S. Allison} (2013): \enquote{Some new
  results on the empirical copula estimator with applications,}
  \emph{Statistics \& Probability Letters}, 83, 1731--1739.

\end{thebibliography}
